\documentclass[useAMS, usenatbib]{mn2e}
\usepackage{times}
\usepackage{graphicx}
\usepackage{multirow}
\begin{document}

\title[Faint discs at polluted white dwarfs]{Signs of a faint disc population at polluted white dwarfs}
\author[C. Bergfors et al.]
{C. Bergfors$^{1}$\thanks{E-mail: c.bergfors@ucl.ac.uk}, J. Farihi$^{1}$\thanks{Ernest Rutherford Fellow}, P. Dufour $^{2}$ and  M. Rocchetto $^{1}$\\
$^{1}$University College London, Department of Physics and Astronomy, Gower Street, London WC1E 6BT, UK\\
$^{2}$D\'{e}partement de Physique, Universit\'{e} de Montr\'{e}al, Montr\'{e}al, QC H3C 3J7, Canada}
\date{Accepted 2014 July 31. Received 2014 July 29; in original form 2014 June 5}

\pagerange{\pageref{firstpage}--\pageref{lastpage}} \pubyear{2014}

\maketitle

\label{firstpage}

\begin{abstract}
Observations of atmospheric metals and dust discs around white dwarfs provide important clues to the fate of terrestrial planetary systems around intermediate mass stars. We present \textit{Spitzer} Infrared Array Camera observations of 15 metal polluted white dwarfs to investigate the occurrence and physical properties of circumstellar dust created by the disruption of planetary bodies.
We find subtle infrared excess emission consistent with warm dust around KUV\,15519+1730 and HS\,2132+0941, and weaker excess around the DZ white dwarf G245-58, which, if real, makes it the coolest white dwarf known to exhibit a $3.6\mu$m excess and the first DZ star with a bright disc. All together our data corroborate a picture where (1) discs at metal-enriched white dwarfs are commonplace and most escape detection in the infrared (possibly as narrow rings), (2) the discs are long lived, having lifetimes on the order of $10^6$\,yr or longer, and (3) the frequency of bright, infrared detectable discs decreases with age, on a timescale of roughly 500 Myr, suggesting large planetesimal disruptions decline on this same timescale.

\end{abstract}


\section{Introduction}

Planetary systems around evolved stars and white dwarfs provide keys to understanding the history of planet formation in our Galaxy, and their fate -- including that of our Solar System.
Although no planetary mass objects have yet been discovered around single white dwarfs, several planets are known to orbit evolved stars and stellar remnants, e.g. pulsars \citep{WolszczanFrail1992}, binary white dwarf + neutron star systems \citep[e.g.][]{Sigurdsson2003}, giants and subgiants 
\citep[e.g.][]{Frink2002, Johnson2007, Sato2008, Bowler2010}. 
These discoveries, together with the fact that more than 95\% of all stars are destined to become white dwarfs, imply that at least some planetary systems survive post-main sequence evolution.
Based on the prevalence of white dwarfs that have metal polluted atmospheres deriving from the accretion of planetary material, at least $\approx30$ per cent still have rocky minor (or major) planets (\citealt{Sion2009}; \citealt{Zuckerman2010}; \citealt*{Koester2014}).

The presence and abundance pattern of metals in the atmospheres of white dwarfs, combined with often-observed infrared excesses, imply accretion of circumstellar material created by the disruption of minor planets within the stellar Roche limit.
Since the first observation of infrared excess from a circumstellar disc around a (single) white dwarf \citep[G29-38;][]{ZuckermanBecklin1987}, clues to the origin of the circumstellar dust have accumulated, primarily from spectroscopic observations. Most white dwarfs have thin atmospheres consisting of pure hydrogen (DA type) or helium (non-DA)\footnote{White dwarfs are classified as D (degenerate) followed by identified optical or ultraviolet spectral features; in our sample we have types A (hydrogen) and B (helium), with Z denoting metal lines. DZ stars have helium-rich atmospheres but are too cool for He\,I absorption to be visible in their spectra \citep{McCookSion1999}.}. Owing to their high surface gravity, the time-scale for any metals to sink out of the atmosphere is short compared to the evolutionary time-scale for white dwarfs cooler than $\rm T_{eff}\la25\,000$\,K, for which radiative forces are negligible \citep{Koester2009}. The metal diffusion time-scale varies from only a few days for hot and young DA white dwarfs to a few $10^6$ yr for helium and old hydrogen white dwarfs. Any heavy element enrichment in the atmospheres of cool white dwarfs therefore implies relatively recent accretion.

The composition of the accreted material provides vital clues to its source. While early discoveries of polluted white dwarfs \citep[e.g.][]{Lacombe1983, Zuckerman2003} could not exclusively determine the origin of the accreted material -- interstellar clouds, comets or asteroids have been suggested -- the evidence has since become compelling that the accretion of planetary debris is the cause of the photospheric pollution and the only source that can adequately explain different observational properties. For instance,
if accretion from the interstellar medium were responsible for the pollution 
it must have occurred relatively recently for DA stars. This theory is strongly disfavoured due to the
lack of nearby dense interstellar clouds, and the relative absence of hydrogen in helium
white dwarfs \citep[see e.g.][]{Farihi2010a}.

The strongest support for accretion of disrupted planetesimals comes from the inferred composition of the circumstellar material, which has been shown to be silicate rich and carbon poor -- in contrast to the interstellar medium which is rich in volatiles such as carbon.
The first infrared spectra with \textit{Spitzer} of the dusty white dwarfs G29-38 and GD\,362 \citep{Reach2005a,Jura2007a} showed strong silicate emission at $9-11\mu$m consistent in shape with zodiacal dust emission, and no signs of polycyclic aromatic hydrocarbon (PAH) emission which is abundant in the interstellar medium at mid-infrared wavelengths.
These features in the infrared spectra have since been confirmed for all white dwarfs with circumstellar discs that have been observed with the $Spitzer$ Infrared Spectrograph \citep*{Jura2009}, and additional silicate emission features were also found at $18-20\mu$m for G29-38 \citep{Reach2009}. 

The metal abundances of the stars mirror the circumstellar material. Analysis of the heavily polluted atmosphere of GD\,362 revealed 15 heavy elements, the relative abundances of which indicated that the debris had a composition that was similar to that of the Earth-Moon system \citep{Zuckerman2007}. 
In total 19 distinct metals have been detected in the atmospheres among several polluted white dwarfs from high resolution spectroscopic observations \citep[e.g.][]{Zuckerman2007,  Melis2011, Dufour2012, KawkaVennes2012, Xu2014}.

To date, 30 previously published metal polluted white dwarfs show excess infrared emission, and in some cases metallic gas emission (\citealt{Gaensicke2006, Gaensicke2008}; \citealt*{Gaensicke2007}; \citealt{Farihi2012a, Melis2012}), indicative of circumstellar debris. These discs are in general similar -- the observed infrared excesses imply emission from hot circumstellar dust 
and a lack of cool dust outside of the stellar Roche limit ($r\approx1R_{\sun}$).
The now standard model for formation of these circumstellar discs 
involves a (as of yet undetected) planet perturbing a minor planetary body into a highly eccentric orbit (\citealt{DebesSigurdsson2002};  \citealt*{Bonsor2011}; \citealt*{Debes2012})
which is tidally disrupted when it comes within the Roche limit. 
The planetary material is viscously spread into a
vertically thin, optically thick flat disc of temperature $T\sim1000$\,K with an inner edge near a radius where silicate grains would rapidly sublimate 
\citep{Jura2003, RafikovGarmilla2012}, from where it is accreted on to the white dwarf, causing the observed metal pollution.
\textit{The composition of the accreted material therefore provides a way of measuring the bulk compositions of the disrupted minor or even major planets}.

\section{Scientific motivation}

To date, more than 200 white dwarfs have been searched for infrared excess, however dust emission from circumstellar discs has only been detected in those that have metal polluted atmospheres \citep{vonHippel2007, Mullally2007, Farihi2012a}. Previous studies have shown that while a significant fraction of polluted white dwarfs show infrared excess emission from circumstellar dust, the majority have no observable discs \citep*{Debes2007, Farihi2009}. 
One possibility for the lack of observed disc emission around metal polluted white dwarfs is that the material is mainly gaseous from high speed collisions in discs with low particle densities \citep*{Jura2007b}. Alternatively, for helium white dwarfs which have long diffusion time, it is possible that the metals detected in the photosphere have been accreted from discs that have since dissipated.

\citet{Jura2008} proposed a model in which massive dust discs, and hence those with observable infrared emission, originate from the disruption of large asteroids, while for the majority of metal polluted white dwarfs continuous accretion of small asteroids creates a mainly gaseous disc with low surface dust density and hence modest infrared excess.
In fact, previous studies have detected relatively subtle excesses around white dwarfs, which can be modelled as narrow dust rings (\citealt*{Farihi2008}; \citealt{Farihi2010b}).

Observationally, the infrared bright disc frequency appears to be correlated with both the accretion rate (averaged over time for helium white dwarfs) and the white dwarf cooling age such that younger white dwarfs with high mass accretion rate are more prone to show infrared excess \citep{Farihi2009, XuJura2012}. This correlation is however somewhat biased since it is observationally more challenging to detect metal pollution in the warmest white dwarfs compared to cooler ones, resulting in a smaller number of relatively young metal polluted stars observed with \textit{Spitzer}.
Discriminating between these two groups of polluted white dwarfs -- those with infrared detectable dust and those without -- could possibly indicate whether the pollutant is one single large asteroid or a collection of several smaller ones \citep{Jura2008}, with implications for the interpretation of the heavy element abundance patters in a spectral analysis.

We observed 15 metal polluted white dwarfs with \textit{Spitzer} IRAC in search for infrared excess indicative of warm circumstellar dust.
The aim of this study is to extend previous surveys by observing additional metal polluted white dwarfs, and especially investigate bright disc frequency among cooler stars.  
In Section 3 we present the \textit{Spitzer} IRAC and complementary ground based observations and data reduction. The results and analysis are presented in Section 4, including the detection of infrared excess consistent with warm circumstellar dust for two, possibly three, of our targets: KUV\,15519+1730, G245-58 and HS\,2132+0941. We discuss the properties of the newly  discovered infrared excesses in the context of disc frequency and lifetime in Section 5, and summarise the results of the survey in Section 6.

\begin{table*}
\caption{List of IRAC imaging targets. Coordinates are measured from the Cycle 7 IRAC images (equinox J2000.0).}
\begin{center}
\begin{tabular}{l r r r r r r l}
\hline\hline
WD	&	Name	&	R. A.	&	Dec	&	SpT 	&	$T_{\rm eff}$	&	[Ca/H(e)]	& Ref. \\
		&		& (h m s)	& ($\degr$\ $\prime$\ $\prime\prime$	)	&		&	($K$)			&			& 	\\
\hline
0122--227	&	KUV\,01223--2245	&	01 24 44.8	&	$-$22 29 07	&	DZ		&	10\,000	&	$-9.4$		&	1	\\
0246+734	&	G245-58			&	02 51 51.8	&	+73 41 34		&	DZ		&	7500		&	$-9.7$		&	2	\\
0543+579	&	GD 290			&	05 47 56.5	&	+57 59 22		&	DAZ		&	8140		&	$-10.3$		&	3	\\
0625+100	&	G105-B2B			&	06 27 37.6	&	+10 02 13		&	DZ		&	8800		&	$-9.5	$		&	4	\\
0816--310	&	SCR\,0818ˆ$-$3110	&	08 18 40.5	&	+31 10 29		&	DZ		&	6460		&	$-9.2$  		&	5	\\
0840--136	&	LP\,786-1			&	08 42 48.3	&	$-$13 47 13	&	DZ		&	4870		 &	$-10.9$		&	5	\\
1009--184	&	WT\,1759			&	10 12 01.5	&	$-$18 43 33	&	DZ		&	6040		&	$-10.6$ 		&	5	\\
1055--039	&	G163-28			&	10 57 47.8	&	$-$04 13 32	&	DZ		&	6500		&	$-10.7$		&	4	\\
1338--311	&	CE\,354			&	13 41 25.7	&	$-$31 24 50	&	DZ		&	8210		&	$-10.0$		&	6	\\
1425+540	&	G200-40			&	14 27 35.8	&	+53 48 30		&	DBAZ	&	14\,490	&	$-9.3	$		&	7, 8	\\
1551+175	&	KUV\,15519+1730	&	15 54 09.1	&	+17 21 24		&	DBAZ	&	15\,550	&	$-6.5	$		&	8	\\
1821--131	&	GJ\,2135			&	18 24 04.4	&	$-$13 08 49	&	DAZ		&	7030		&	$-10.7$		&	3	\\
2132+096	&	HS\,2132+0941		&	21 34 50.8	&	+09 55 20		&	DAZ		&	13\,200	&	$-7.7	$		&	9	\\
2157--574	&	WT\,853			&	22 00 45.1	&	$-$57 11 24	&	DAZ		&	7220		&	$-8.0	$		&	2, 10	\\
2230--125	&	HE\,2230$-$1230	&	22 33 38.8	&	$-$12 15 28	&	DAZ		&	20\,300	&	$-6.3$		&	9	\\

\hline
\label{targets}
\end{tabular}
\end{center}
\begin{flushleft}
References: (1) \citet{Friedrich2000}; 
(2) this paper;
(3) \citet{Zuckerman2003};
(4) \citet*{Dupuis1993};
(5) \citet*{Giammichele2012};
(6) \citet{Dufour2007}; 
(7) \citet{Kenyon1988};
(8) \citet{Bergeron2011};
(9) \citet{Koester2005};
(10) \citet{Subasavage2007}
\end{flushleft}
\end{table*}

\section{Observations and data reduction}
\subsection{Target selection and $Spitzer$ IRAC observations}
Our 15 targets were selected in 2010 as the bulk of metal enhanced nearby and relatively bright white dwarfs that had not been observed with the $Spitzer$ Infrared Array Camera (IRAC). The majority of the targets are moderately polluted with [Ca/H(e)]$\la-7.5$ and mainly with effective temperatures $T_{\rm eff}\le10\,000$\,K. These targets belong to a region of temperature and Ca abundance phase space that had not been well explored in previous studies. Three targets, KUV\,15519+1730, HS\,2132+0941 and HE\,2230--1230, are warm ($T_{\rm eff}\ga13\,000$\,K) and highly metal polluted ([Ca/H(e)]$\ga-7.5$) white dwarfs, thus likely to have observable excess infrared emission from circumstellar dust. 

Warm \textit{Spitzer} \citep{Werner2004} infrared observations were carried out with IRAC \citep{Fazio2004a} in Cycle 7 within programs 70037 and 70116.
30 medium size dithers of 30\,s integrations were obtained in the cycling pattern for each target in IRAC channels 1 ($3.6\mu$m) and 2 ($4.5\mu$m).  When combined to a 900\,s integration mosaic the many dithers efficiently reduce artefacts such as cosmic rays and bad pixels. Table \ref{targets} lists the observed targets, their coordinates at the epoch of observation, spectral types, published temperatures and calcium abundances.

\subsection{$Spitzer$ IRAC photometry}

The IRAC basic calibrated data (BCD) reduction pipeline produces flux-calibrated data frames from which photometry with a few per cent accuracy can be performed either directly on the individual images or after mosaic combination. We combined the additional artefact mitigated, corrected BCD (CBCD) images to a create a co-added mosaic image using the \begin{small}MOPEX\end{small} post-BCD software\footnote{Details on the reduction pipeline and additional software can be found in the IRAC Instrument Handbook and related documentation at the $Spitzer$ Science Center (SSC) website at http://irsa.ipac.caltech.edu/data/SPITZER/docs/irac/} with a pixel size 0.6 arcsec pixel$^{-1}$.

We perform photometry following \citet{Farihi2008}.
The main effects that contribute to increase photometric uncertainties arise from the array location of the target, and the pixel phase effect where the position of the point-spread function (PSF) centroid on the peak pixel affects the photometric result \citep{Reach2005b, Hora2008}. Both the array-location and pixel-phase effects are expected to average out for well dithered data such as ours. Colour corrections are assumed to be negligible.

Aperture and PSF-fitting photometry was performed on the mosaic combined images for each target and filter using the \begin{small}IRAF APPHOT\end{small} and \begin{small}DAOPHOT\end{small} packages. To avoid photometric confusion, we perform photometry either using a small aperture of radius $r=4$ pixel and 24--40 pixel sky annulus, 
with the corrections recommended in the IRAC Instrument Handbook,
or PSF-fitting on the mosaic combined images. 
We calculate $1\sigma$ errors of the aperture photometry by measuring the standard deviation in the sky, and adopt the errors reported by \begin{small}DAOPHOT\end{small} for the PSF-fitting photometry, adding in quadrature a 5 per cent absolute calibration error to each measurement \citep[see][]{Farihi2008}.
Additional photometry was performed for a few targets using the Astronomical Point-source Extractor (\begin{small}APEX\end{small}) in \begin{small}MOPEX\end{small} to deblend some targets or confirm an infrared excess (see Section 3.4). The \textit{Spitzer} IRAC photometry is listed in Table \ref{photometry}.

\begin{figure*}
 \centering
  \begin{tabular}{c c}
 \includegraphics[width=80mm]{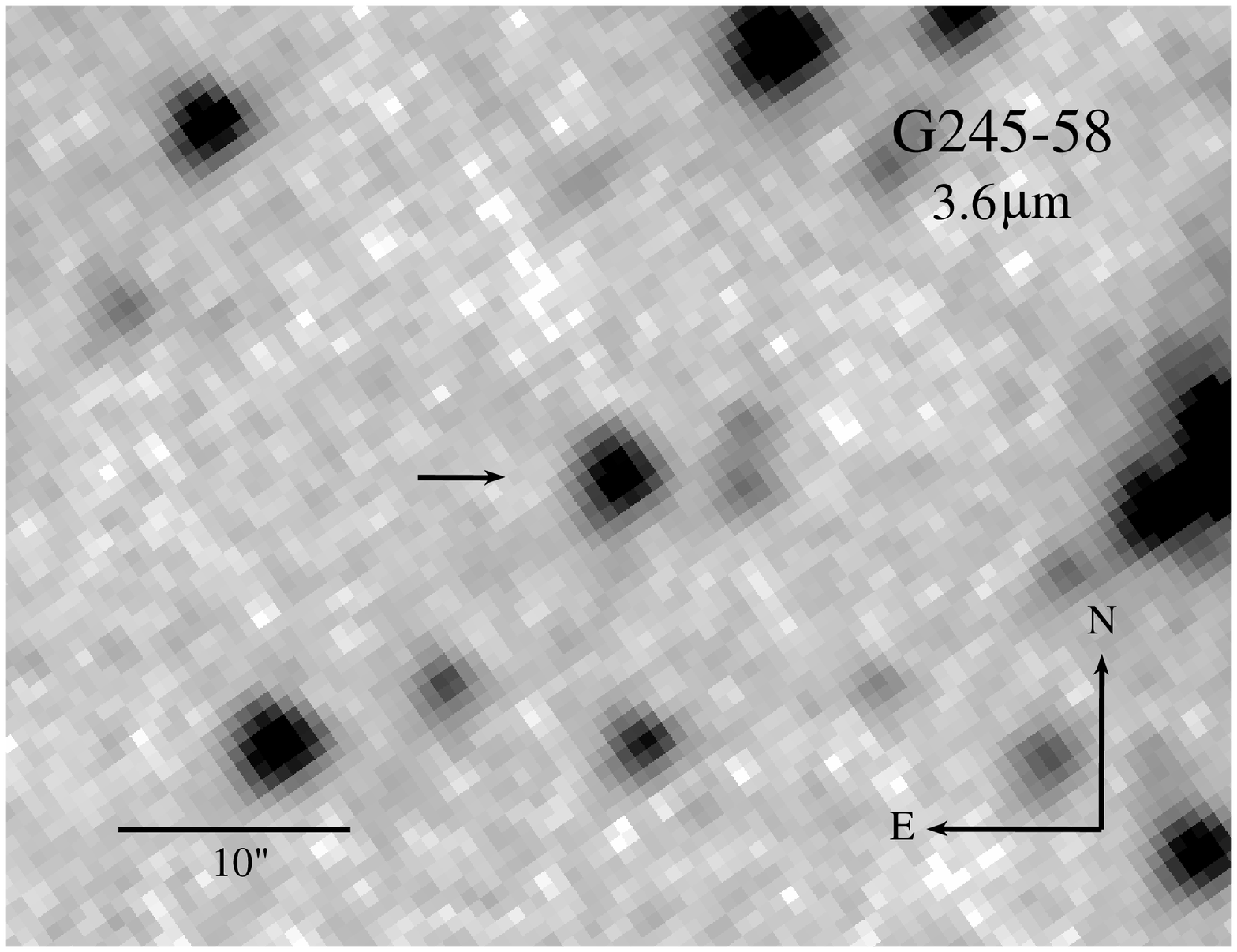} &  \includegraphics[width=80mm]{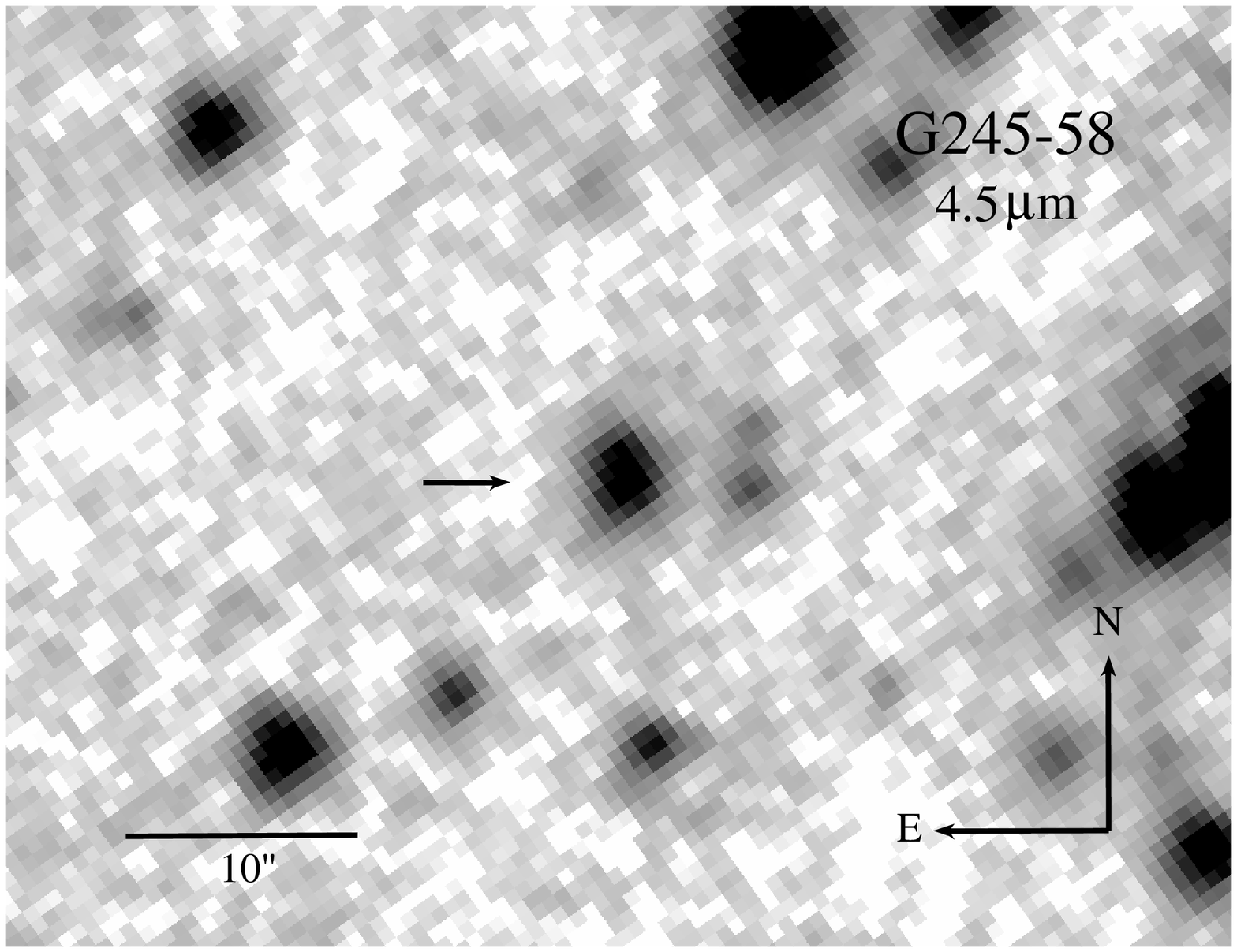}\\
 \end{tabular}
    \caption{\textit{Spitzer} IRAC 0\farcs6 pixel$^{-1}$ images in $3.6\mu$m (Ch 1) and $4.5\mu$m (Ch 2) of G245--58. The images are shown on identical square root scales. The target star appears slightly elongated at $4.5\mu$m, most likely due to a background object.}
\label{WD0246_Ch1Ch2}
 \end{figure*}

\begin{table}
\caption{IRAC Ch1 and Ch2 photometry. }
\begin{center}
\begin{tabular}{l r r  l}
\hline\hline
WD 	&	$F_{3.6\mu m}$\	&	$F_{4.5\mu m}$	&	IR excess	\\
		&  ($\mu$Jy)	&	($\mu$Jy)	&	\\
\hline	
0122--227	&	$52\pm3$	&	$38\pm2$	&	N	\\
0246+734$^a$&	$65\pm5$	&	$47\pm5$	&	Y	\\
0543+579	&	$172\pm9$	&	$106\pm6$	&	N	\\
0625+100	&	$82\pm8$	&	$56\pm7$	&	N	\\
0816--310	&	$382\pm19$	&	$248\pm13$	&	N	\\
0840--136	&	$550\pm28$	&	$366\pm18$	&	N	\\
1009--184	&	$474\pm24$	&	$291\pm15$	&	N	\\
1055--039	&	$129\pm7$	&	$88\pm5$	&	N	\\
1338--311	&	$48\pm5$	&	$32\pm4$	&	N	\\
1425+540	&	$163\pm8$	&	$102\pm5$	&	N	\\
1551+175	&	$29\pm2$	&	$26\pm2$	&	Y	\\
1821--131$^b$	&		- 	&		- 	&	-	\\
2132+096$^a$	&	$98\pm5$	&	$79\pm5$	&	Y	\\
2157--574	&	$252\pm13$	&	$152\pm8$	&	N	\\
2230--125	&	$53\pm3$	&	$34\pm2$	&	N	\\
\hline
\label{photometry}

\end{tabular}
\end{center}
\begin{flushleft}
$^a$ Weighted average of fluxes extracted with DAOPHOT and APEX (see Section 3.4).\\
$^b$ Accurate fluxes could not be extracted for this target due to severe confusion.
\end{flushleft}
\end{table}

\subsection{Ground based near-infrared photometry}

Additional ground-based observations were obtained for some $Spitzer$ IRAC targets in the near-infrared  $JHK_s$ bands with the Long-slit Intermediate Resolution Infrared Spectrograph  \citep[LIRIS; ][]{Manchado1998} at the 4.2\,m William Herschel Telescope (WHT) in 2011 March and October, and with Son of ISAAC \citep*[SOFI; ][]{Moorwood1998} at the 3.6\,m New Technology Telescope (NTT) in 2011 August and 2012 January. The science targets were observed in a continuous dither pattern with individual exposure times of 30, 15, and 10\,s at $J$, $H$, and $K_s$ respectively for a total integration time of 270\,s in each filter. Each night three standard star fields from the ARNICA catalogue \citep{Hunt1998} were observed for flux calibration.

Standard data reduction was performed. The median sky subtracted frames were shifted and average combined into a single image on which aperture photometry with IRAF \begin{small}APPHOT\end{small} or, for targets with nearby sources, \begin{small}DAOPHOT\end{small}, was performed using standard aperture radii between $3.8$ and $4.3$ arcsec and sky annuli in the range $5.0-8.6$ arcsec. Smaller apertures were used for the faintest sources ($m>15.5$\,mag), with corrections derived from bright stars in the same image. The flux calibration was good to 5 per cent or better on each night.

\subsection{Photometric extraction of blended sources}
G245-58 appears slightly elongated at $4.5\mu$m and therefore blending with a background object was suspected (Fig. \ref{WD0246_Ch1Ch2}).
The additional source is brighter at $4.5\mu$m, where it appears relatively extended and diffuse, and hence it is likely a background galaxy.

In order to avoid contamination from the neighbouring source we performed PSF-fitting photometry on the science target using both \begin{small}DAOPHOT\end{small} and \begin{small}APEX\end{small}, in addition to aperture photometry, with the results listed in Table \ref{Table_daophot_apex}.
A comparison between the aperture and PSF-fitting photometry showed that while the aperture photometry gives slightly higher flux densities, the values are consistent with those from the PSF-fitting photometry within errors. 
We also performed aperture photometry on the residual images generated by \begin{small}DAOPHOT\end{small} after subtraction of the white dwarf, and found a residual flux density of $2-4$ per cent of the extracted target flux in both channels. This is well within the
error bars and identical to that found for isolated targets analysed in the same manner.
The potential contamination by the background source is therefore much smaller than the total flux excess measured compared to a white dwarf atmospheric model (see Section 4.3) and the infrared excess is thus likely associated with circumstellar dust around the white dwarf.

As an additional test we compared the position relative to other stars in the field in the LIRIS $JHK_s$ and the $3.6$ and  $4.5\mu$m IRAC images, since a shift of the PSF peak could indicate contamination of the photometry by the background source. We find no significant shift and conclude that the flux we extracted and measured in the IRAC images is that of the target. 
The PRF-fitting photometry with \begin{small}APEX\end{small} gives marginally different values, and we adopt a weighted average from the two methods.

Based on our previous experience, approximately 1 out of 20 white dwarfs observed with IRAC needs some photometric deblending, but are typically recoverable \citep[e.g.][]{Farihi2010b}. Our adopted flux measurement for G245-58 at $4.5\mu$m is $47\mu$Jy, corresponding to 16.5 magnitudes (Vega). In order to assess the probability that the flux we measure is that of a background source and not the science target, we use IRAC $4.5\mu$m source counts from \citet{Fazio2004b}.
For the worst case scenario in which the galaxy contributes most of the flux measured at this wavelength, there are less than $2\times10^4$ galaxies deg$^{-2}$ within a 1 mag interval centred at 17.0 mag.
With a photometric aperture radius of $2.4$ arcsec, we find a probability of less than 1 per cent that the measured flux is due to a sufficiently bright background galaxy.

Two of our targets, WD\,0625+100 and WD\,1338--311, suffer from some blending with nearby stars. In both cases PSF-fitting photometry successfully de-convolved both point sources. No infrared excess is observed for these targets in IRAC, but flux contamination is the likely cause of excess seen in the \textit{Wide-field Infrared Survey Explorer} \citep[\textit{WISE}; ][]{Wright2010} $3.4\mu$m flux of WD\,0625+100 (see Fig. \ref{ModelFits2}). We also perform photometry with \begin{small}APEX\end{small} and \begin{small}DAOPHOT\end{small} for HS\,2132+0941 to confirm the infrared excess which is only seen at $4.5\mu$m. The photometry using both methods and the adopted values are listed in Table \ref{Table_daophot_apex}.

\begin{table}
\caption{IRAF and APEX photometry of G245-58 and HS\,2132+0941.}
\centering
\begin{tabular}{l  r r r}
\hline\hline
Wavelength 		&	APEX		&	DAOPHOT	&	Adopted flux\\
($\rm \mu m$)	&	($\rm \mu Jy$)	&	($\rm \mu Jy$)	&	($\rm \mu Jy$) \\
\hline
\multicolumn{4}{c}{G245-58}\\
\hline
$3.6$	&	$65\pm4$	&	$64\pm5$	&	$65\pm5$	\\
$4.5$	&	$45\pm3$	&	$50\pm5$	&	$47\pm5$	\\
\hline
\multicolumn{4}{c}{HS\,2132+0941}\\
\hline
$3.6$	&	$99\pm5$ &	$96\pm6$ &	$98\pm5$ \\
$4.5$	&	$81\pm4$	&	$77\pm5$	&	$79\pm5$ \\
\hline
\label{Table_daophot_apex}
\end{tabular}
\end{table}

\begin{table}
\caption{Ground-based $JH K_s$ photometry. All errors are 5 per cent}
\centering
\begin{tabular}{l r r  r r l}
\hline\hline
WD 	&	$J$	&	$H$	&	$K_s$ & Instrument 	\\
		&  (mag)	&	(mag]) 	&	(mag) & /Telescope   \\
\hline
0122--227	&	16.83 	&	16.82 	&	16.83 	&	SOFI/NTT		\\
0246+734	&	16.93 	&	16.85 	&	16.84 	&	LIRIS/WHT	\\
0625+100	&	16.35 	&	16.32 	&	16.30 	&	LIRIS/WHT	\\
1055--039	&	15.98 	&	15.92 	&	15.84 	&	SOFI/NTT		\\
1338--311	&	17.03 	&	16.99 	&	17.06 	&	SOFI/NTT		\\
1551+175	&	17.52 	&	17.64 	&	17.56 	&	SOFI/NTT		\\
1821--131	&	14.51 	&	14.28 	&	14.21 	&	SOFI/NTT		\\
2132+096	&	16.15 	&	16.17 	&	16.17 	&	SOFI/NTT		\\

\hline
\label{JHKphot}
\end{tabular}
\end{table}

\section{Results and analysis}
\subsection{Spectral Energy Distribution}

\begin{figure*}
\setcounter{figure}{1}
 \centering
 \begin{tabular}{c c}
 \includegraphics[width=69mm, angle=90]{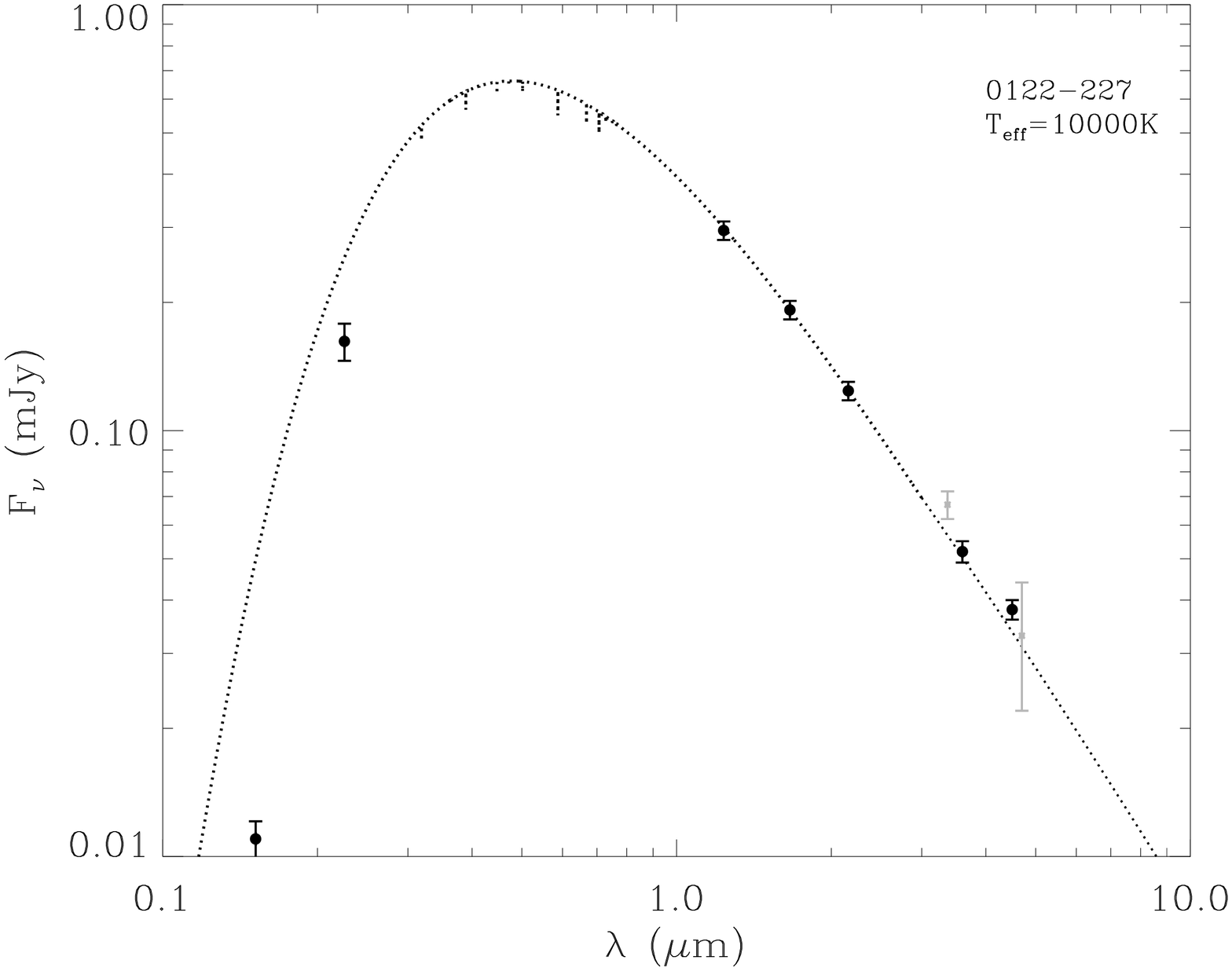} &  \includegraphics[width=69mm, angle=90]{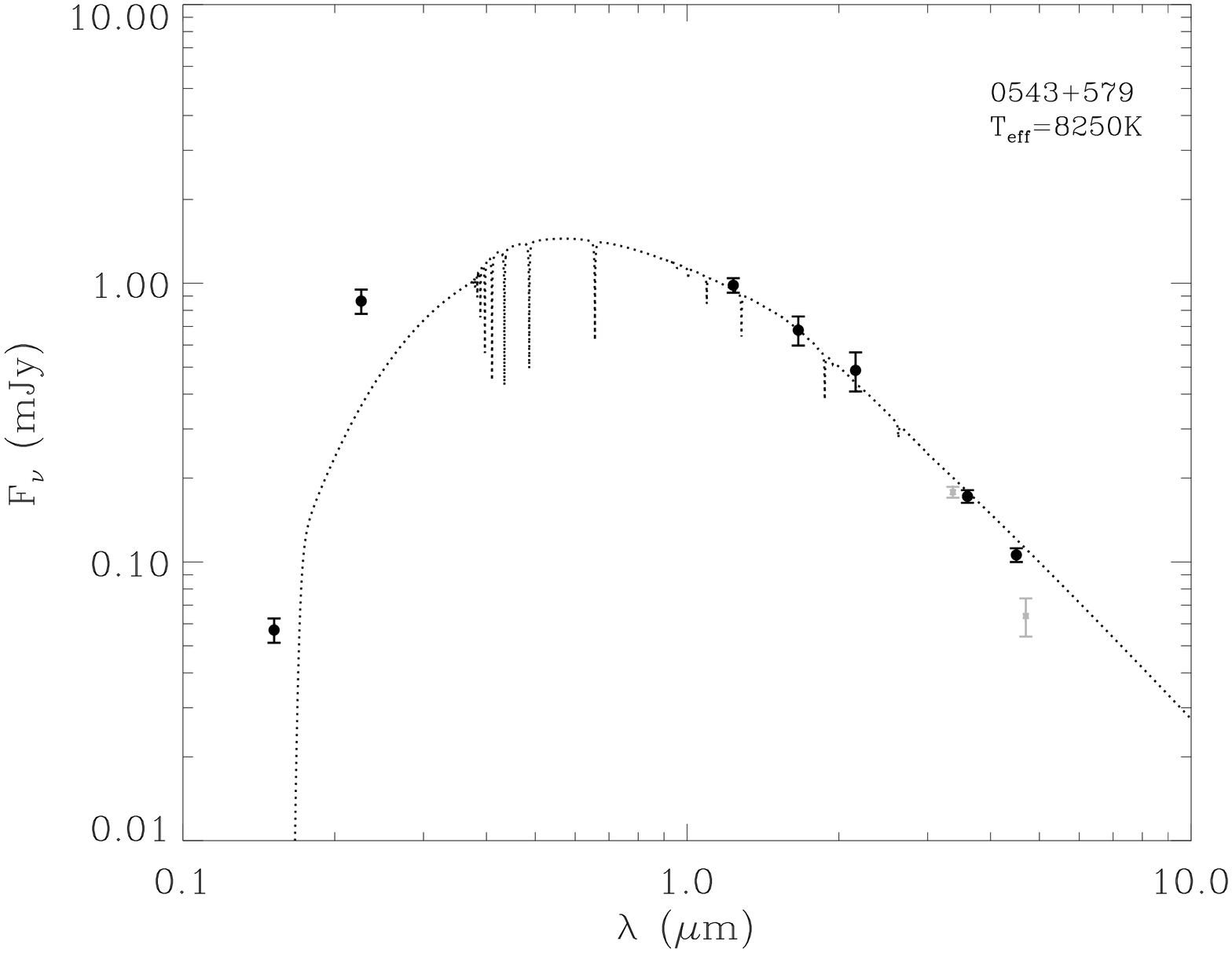}\\
  \includegraphics[width=69mm, angle=90]{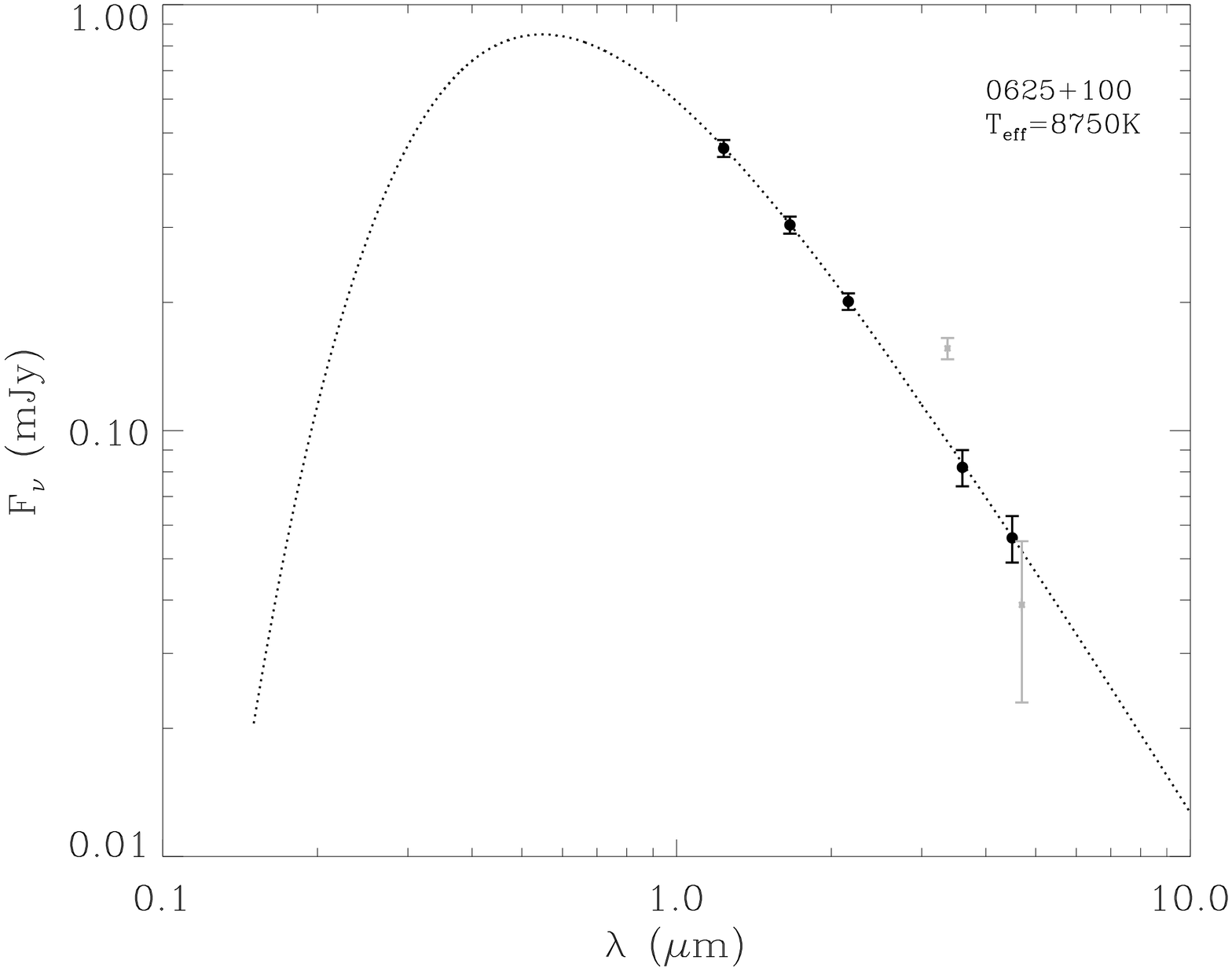} &  \includegraphics[width=69mm, angle=90]{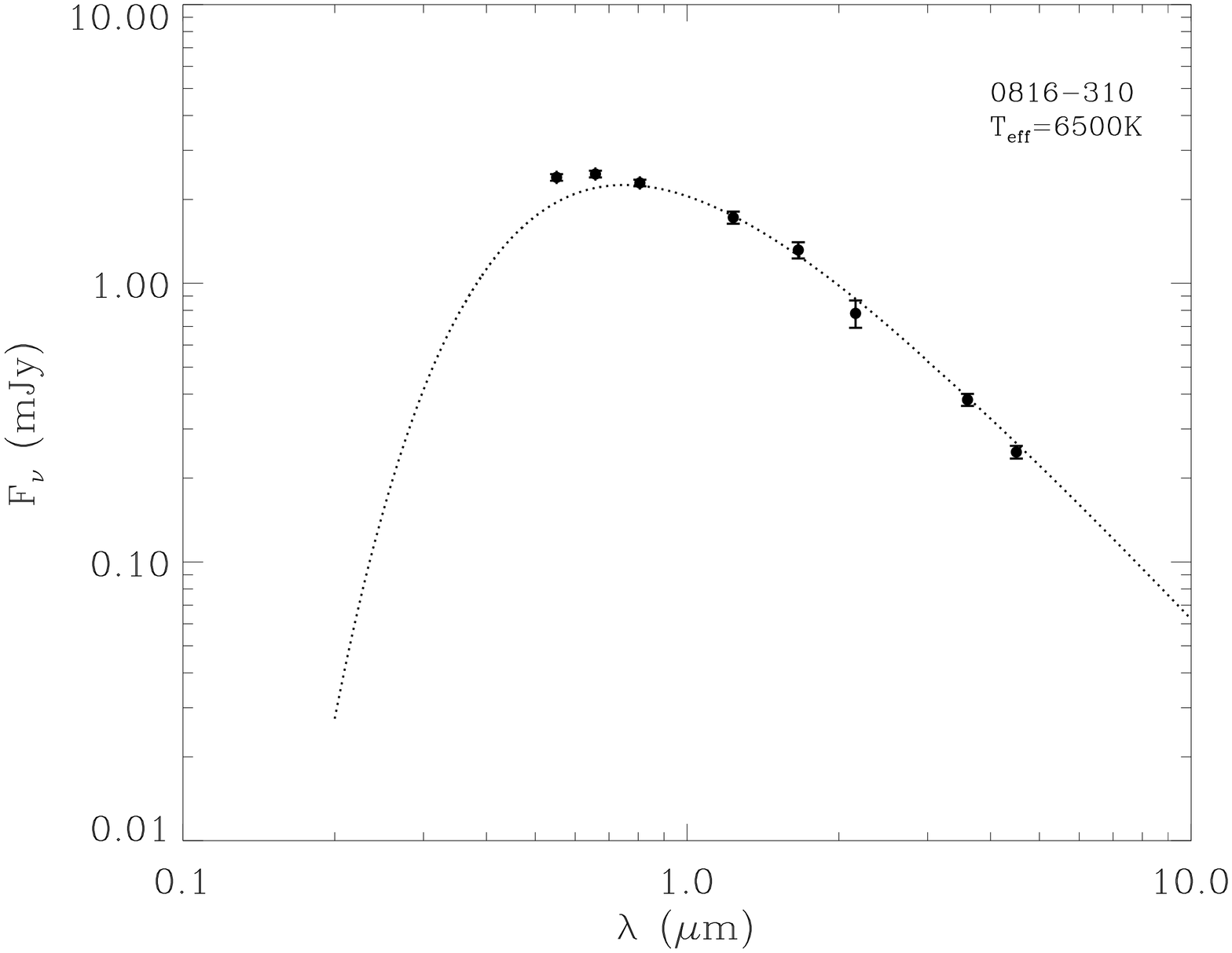}\\
 \end{tabular}
  \caption{Spectral energy distributions for target stars. Photometric data are from \textit{GALEX}, SDSS, \citet{Subasavage2007, Subasavage2008}, 2MASS, Table 4, WISE and IRAC (see Tables \ref{photometry} and \ref{JHKphot}). The \textit{WISE} $4.6\mu$m photometry (light grey) have been slightly offset from the central wavelength in the plot for clarity.}
\label{ModelFits1}
\end{figure*}

\begin{figure*}
\setcounter{figure}{1}

 \centering
 \begin{tabular}{c c}
 \includegraphics[width=69mm, angle=90]{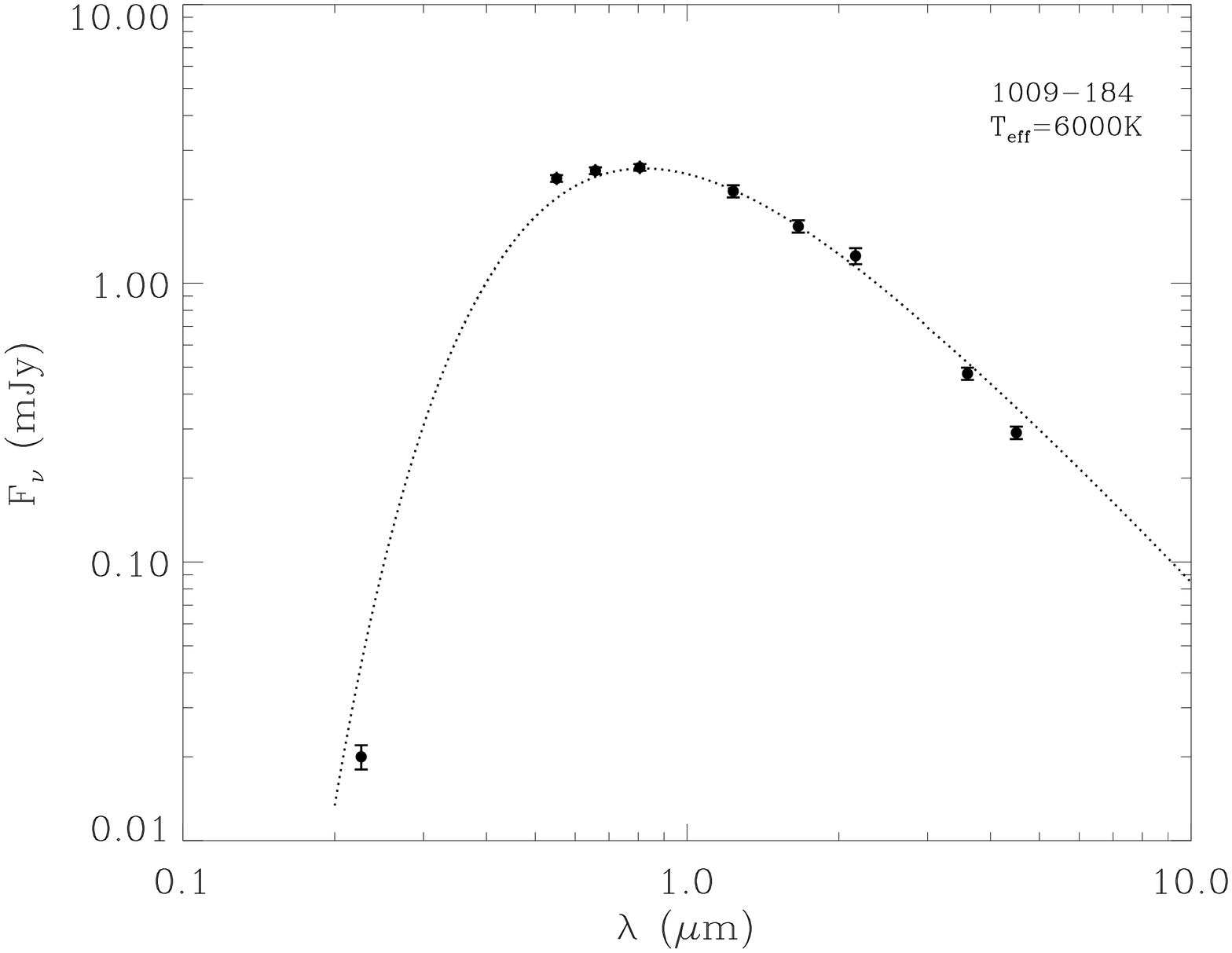} &  \includegraphics[width=69mm, angle=90]{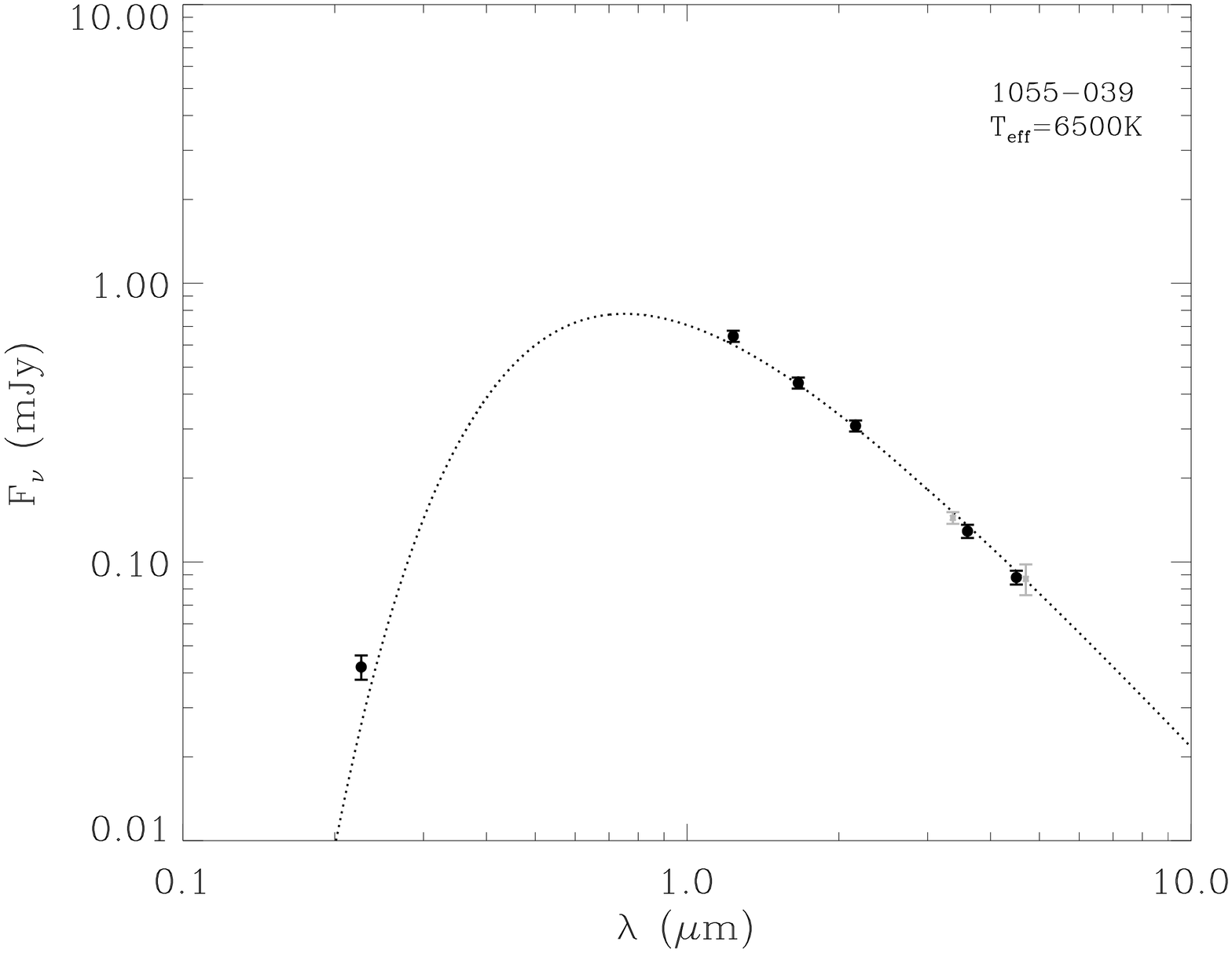}\\
  \includegraphics[width=69mm, angle=90]{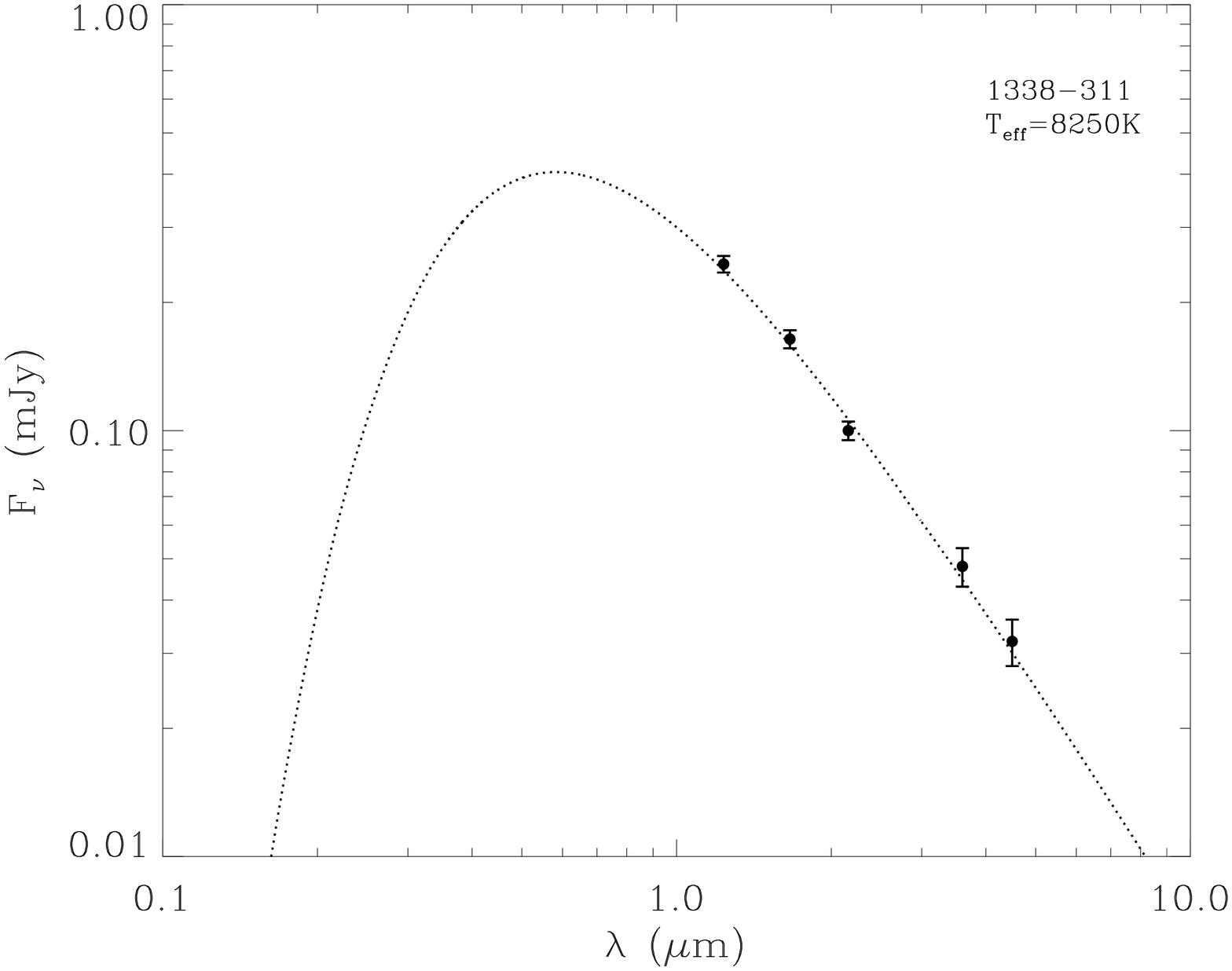} &  \includegraphics[width=69mm, angle=90]{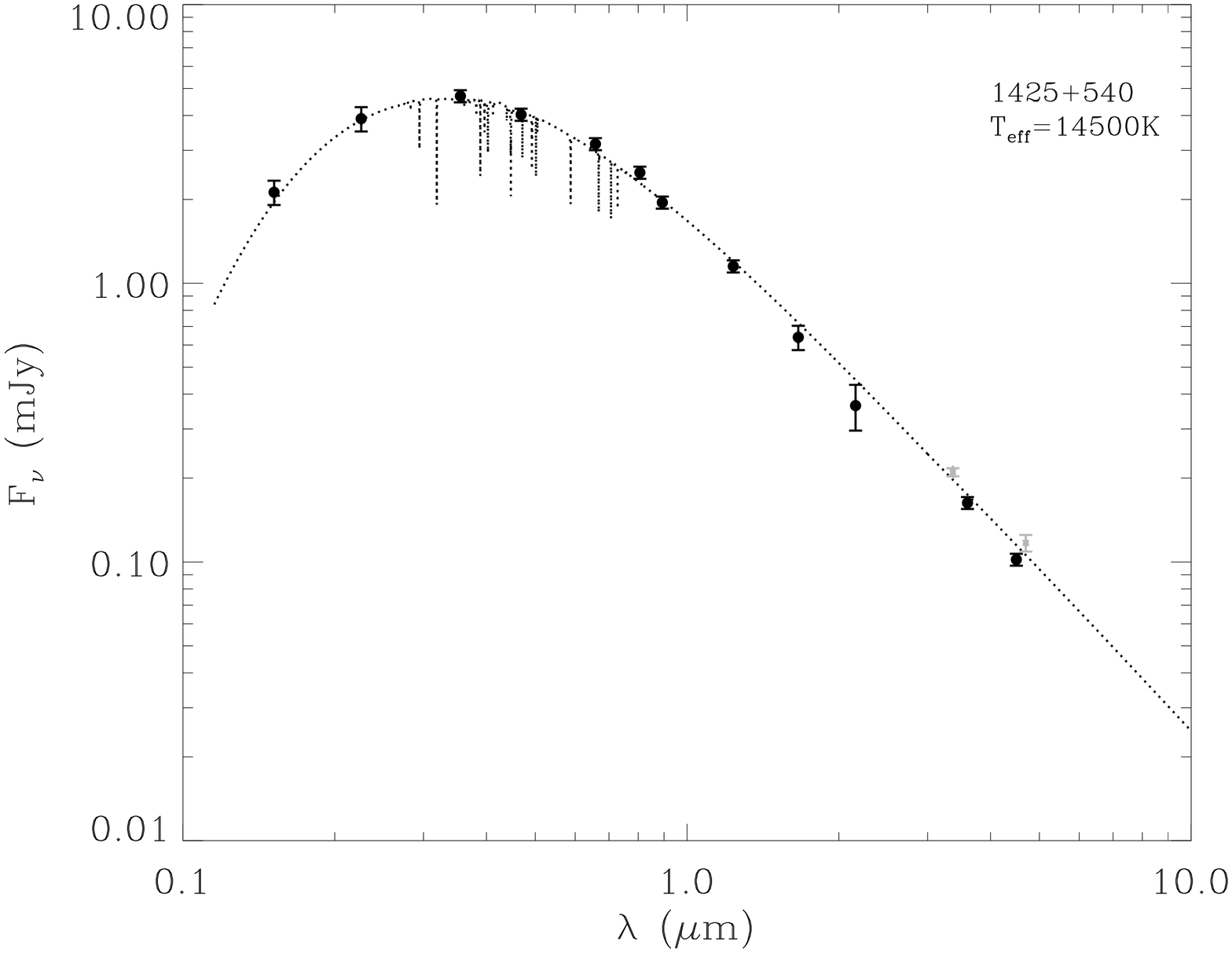}\\
   \includegraphics[width=69mm, angle=90]{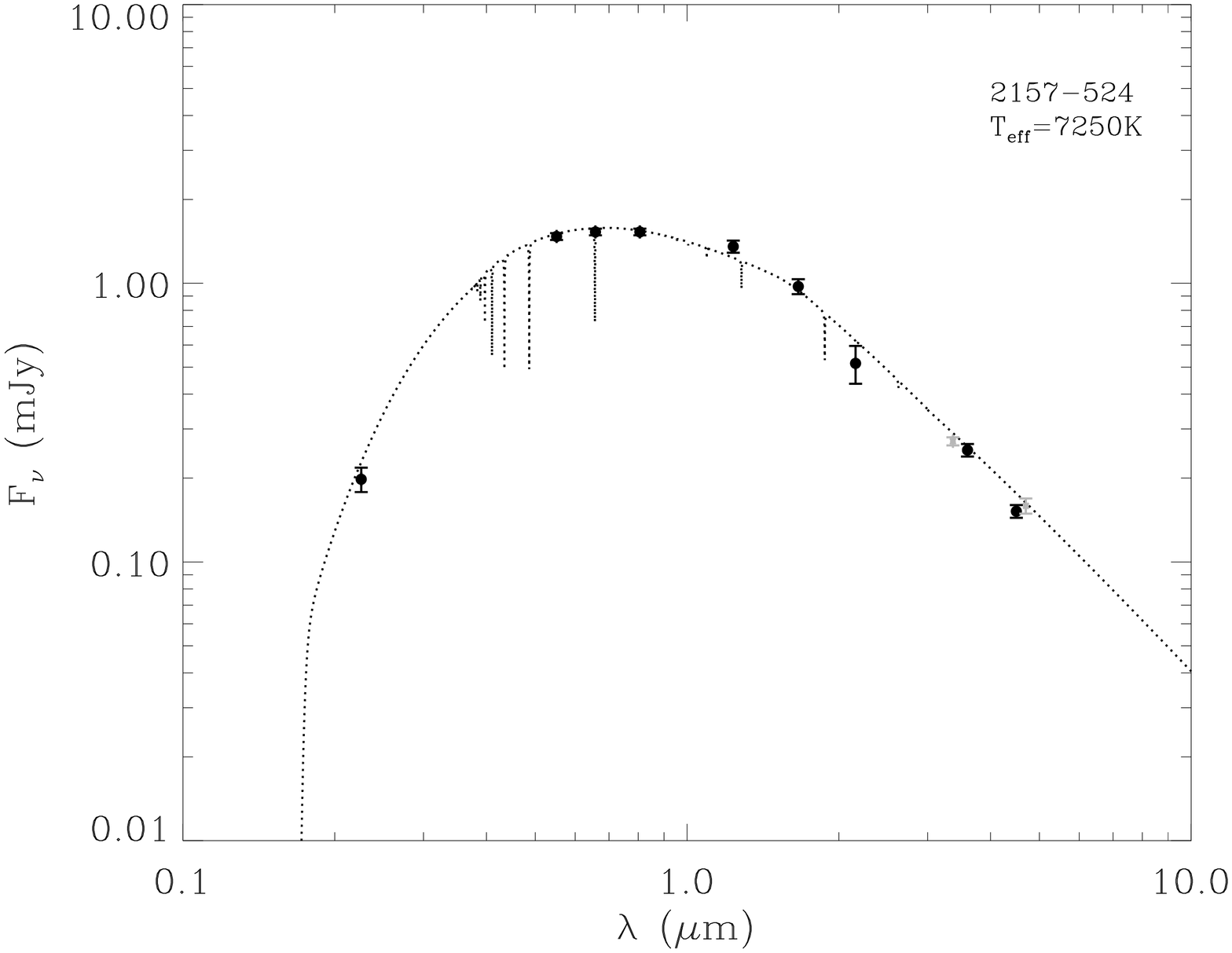} &  \includegraphics[width=69mm, angle=90]{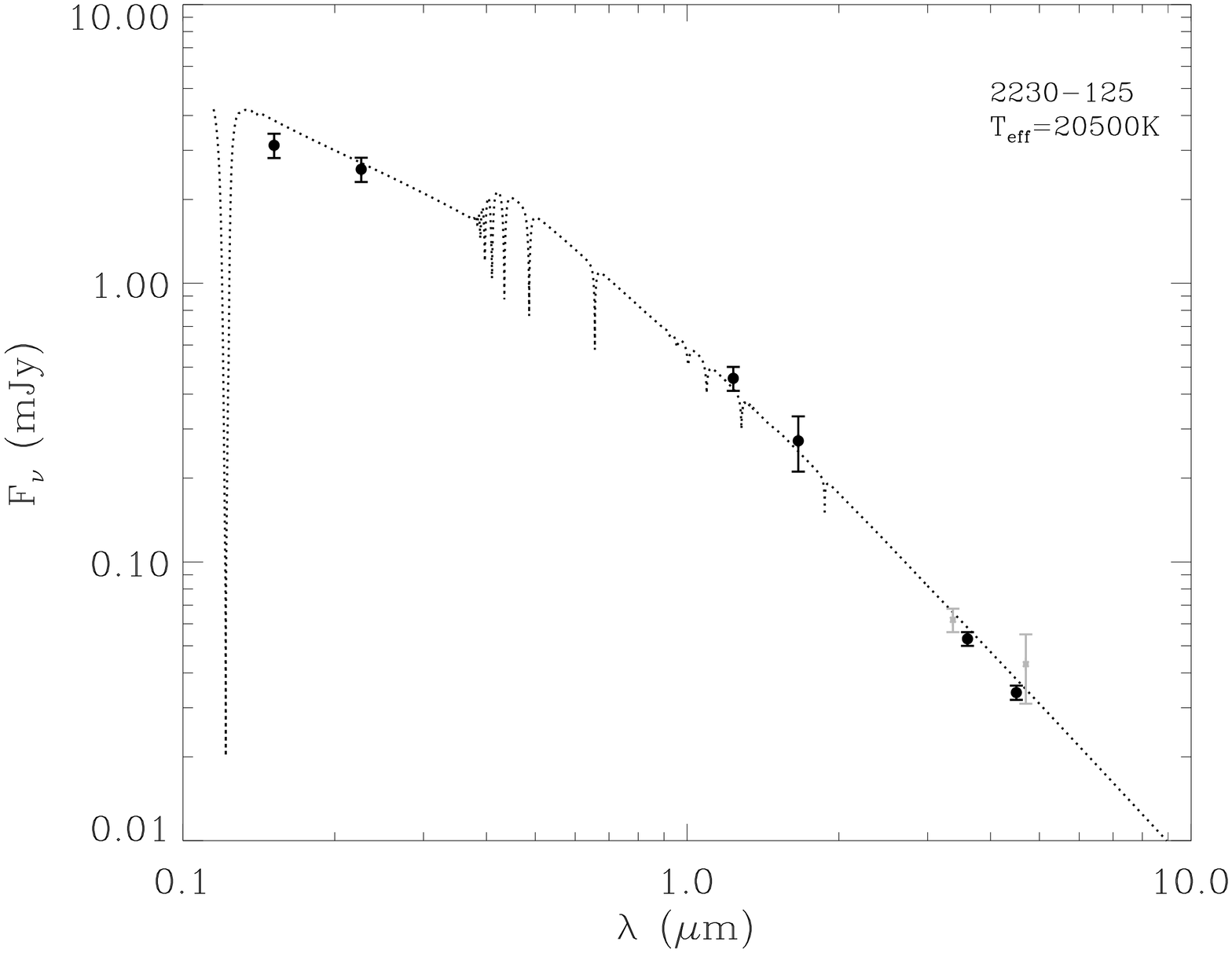}\\
 \end{tabular}
  \caption{\textit{ - continued} }
\label{ModelFits2}
\end{figure*}

\begin{figure}
 \centering
 \includegraphics[width=6.5cm, angle=90]{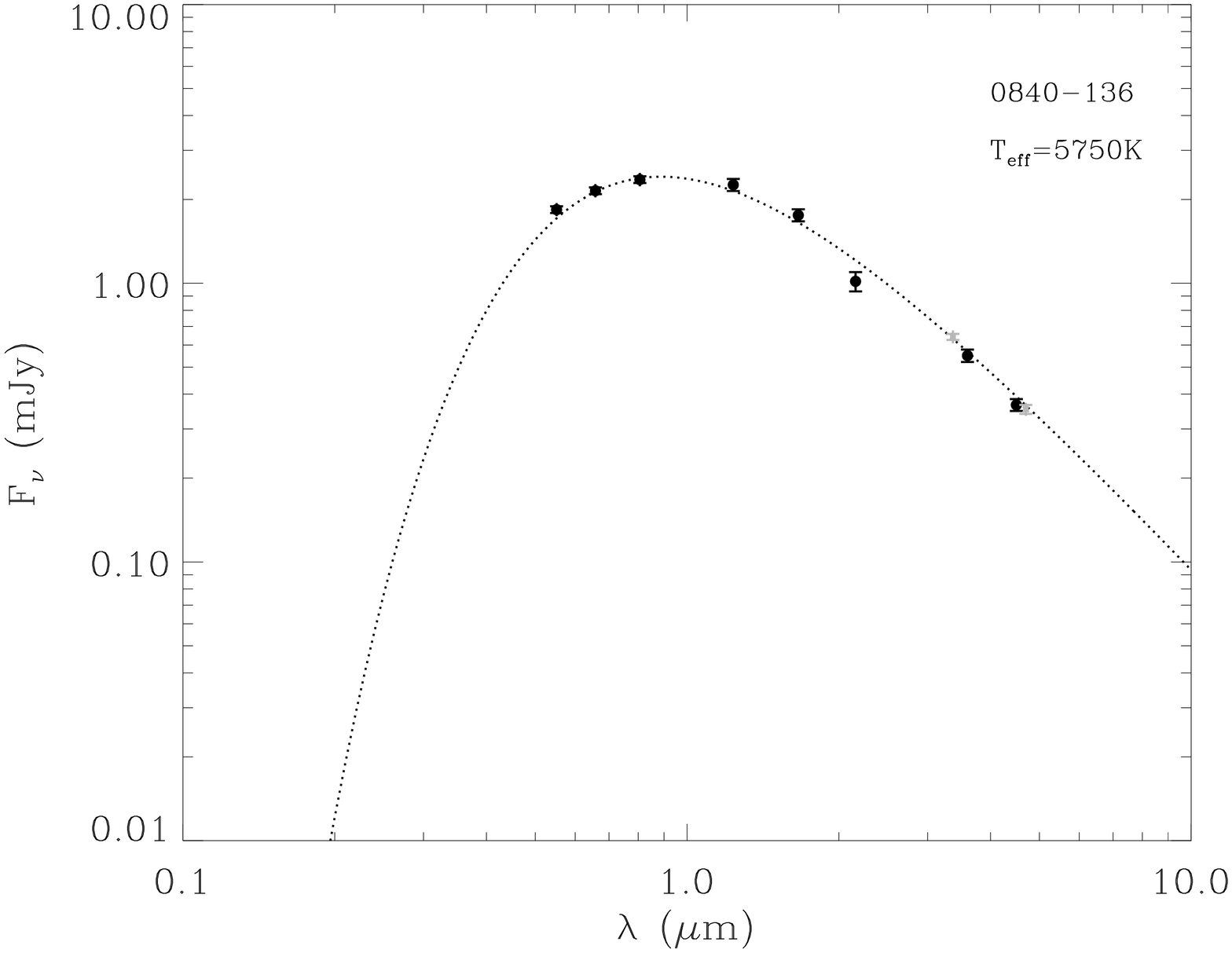}
  \caption{(See Fig. 2 caption). The photometry for WD\,0840--136 is best fitted with a $T=5750$\,K black body (Section 4.5).}
\label{ModelFits3}
\end{figure}

Models of pure hydrogen and helium atmosphere white dwarfs \citep{Koester2009b} for log\,$g=8.0$ 
were scaled to fit available photometric data in the optical from the Sloan Digital Sky Survey \citep[SDSS; ][]{Stoughton2002}, \citet{Subasavage2007,Subasavage2008}, and in the near-infrared from Table \ref{JHKphot} or the Two Micron All Sky Survey \citep[2MASS;][]{Skrutskie2006}.  We also plot in Fig. 2--4 ultraviolet photometry from the \textit{Galaxy Evolution Explorer} \citep[\textit{GALEX}; ][]{Martin2005} if available, with 10 per cent error bars. The far-ultraviolet (FUV) and near-ultraviolet (NUV) data are however not considered for the model fitting as the fluxes may be affected by metal blanketing and extinction \citep{Koester2011}.
When fitting the models we assumed error bars of 5 per cent for the SDSS photometry, errors as listed in catalogue for 2MASS data (minimum 5 per cent), 
and 3 per cent error bars for Johnson \textit{V} and Cousins \textit{R} and \textit{I} photometric data from \citet{Subasavage2007,Subasavage2008}. 
Figs \ref{ModelFits1} and \ref{ModelFits3} show the spectral energy distributions of the targets without infrared excess. Photometry from \textit{WISE} at $3.4$ and $4.6\mu$m is plotted for comparison in these figures if available; it is not however considered for model fitting. \textit{WISE} data are insufficient in spatial resolution and sensitivity for these observational goals (\citealt{Melis2011}; Rocchetto et al., in preparation).

Two of the 15 targets show clear infrared excess compared to the white dwarf atmospheric model at $3.6$ and $4.5\mu$m: KUV\,15519+1730 shows a $4\sigma$ excess at both $3.6$ and $4.5\mu$m, and a $3\sigma$ excess at $2.2\mu$m, compared to the best model fit to the optical and near-infrared data. We detect a less significant ($2.5\sigma\approx20-25$ per cent) excess in both wavelength bands for G245-58 in what is likely the most subtle disc emission detected at at white dwarf. HS\,2132+0941 exhibits a $4\sigma$ excess at $4.5\mu$m compared to the expected photospheric flux, but no significant excess at shorter wavelengths. The spectral energy distributions of these targets are shown in Fig. \ref{excesses}. 

We model the infrared emission of these three stars using opaque and geometrically flat dust rings \citep{Jura2003}, which well describe the thermal dust emission observed at metal-rich white dwarfs. For each star the stellar radius to distance ratio $R_{*}/d$ is calculated using the cooling models of the Montr\'{e}al group\footnote{http://www.astro.umontreal.ca/$\sim$bergeron/CoolingModels/} for the adopted stellar atmospheric model temperature and assuming log\,$g=8.0$ (\citealt{HolbergBergeron2006, KowalskiSaumon2006}; \citealt{Bergeron2011}; \citealt*{Tremblay2011}), and three free parameters are varied to achieve a best fit to the near-infrared and IRAC data: the inner and outer disc temperature, $T_{\rm in}$ and $T_{\rm out}$, and the inclination $i$ of the disc. Corresponding disc radii are calculated following \citet{ChiangGoldreich1997}.

While the inner edge temperature can be relatively well constrained by the emission at shorter wavelengths ($2.2$ and $3.6\mu$m), there is significant degeneracy in these models and they should be considered descriptive rather than definitive. For subtle excesses as found for these three targets, the thermal dust emission can be equally well fitted with relatively wide rings at high inclination, or as narrow rings at modest inclination \citep[][]{Girven2012}. We therefore present two sets of models for each target with excess emission in Table \ref{disctable} and Fig. \ref{excesses}. These model disc parameters do not represent the actual disc properties, but give representative, physically realistic disc temperatures consistent with the data.
Importantly, all realistic models place these discs completely within the Roche limit, and thus consistent with disrupted planetesimals. 
The targets and their infrared excesses are discussed more in detail in the following subsections. 

\subsection{KUV\,15519+1730}
\begin{figure}
 \centering
 \begin{tabular}{c}
 \includegraphics[width=8.5cm]{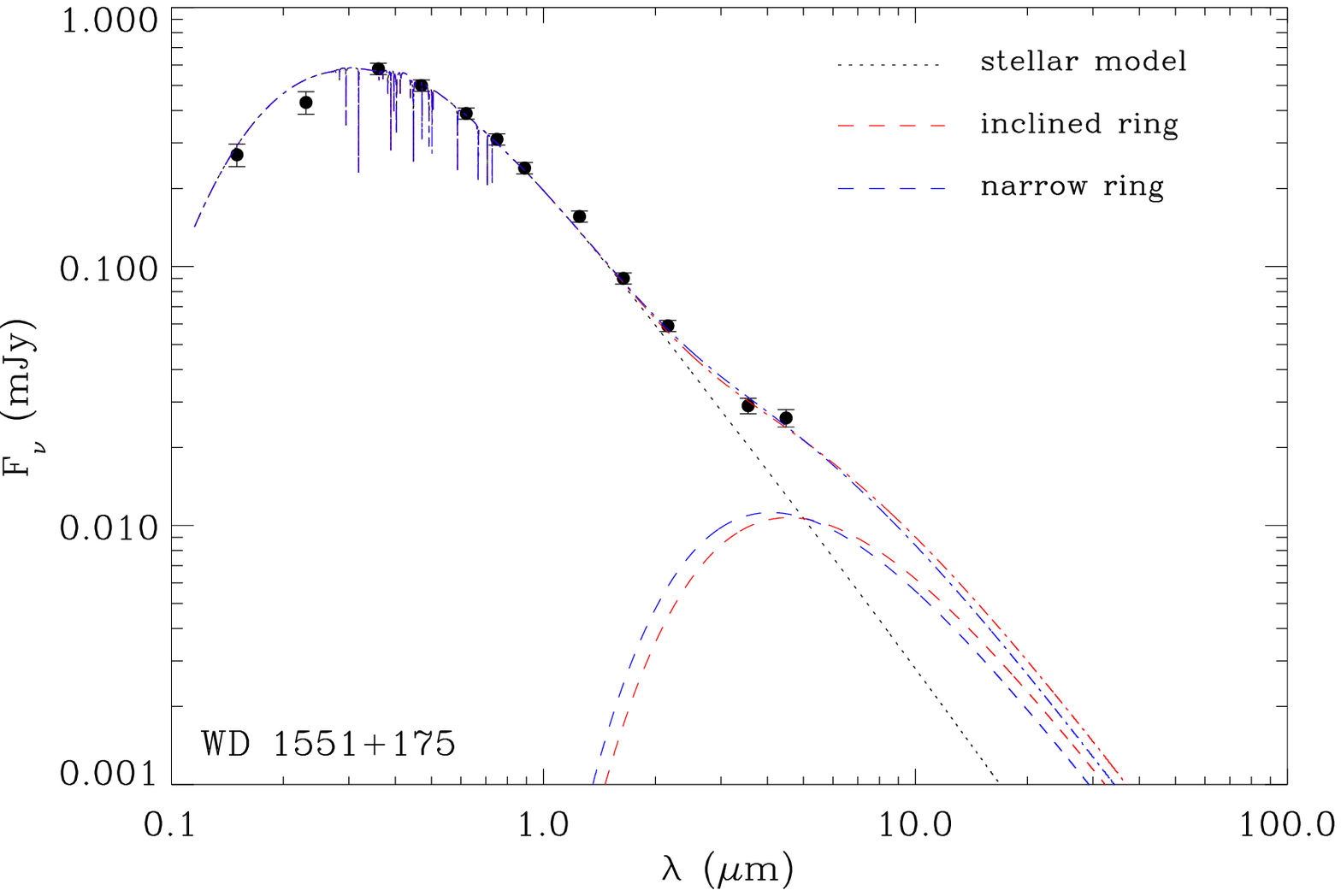}\\
 \includegraphics[width=8.5cm]{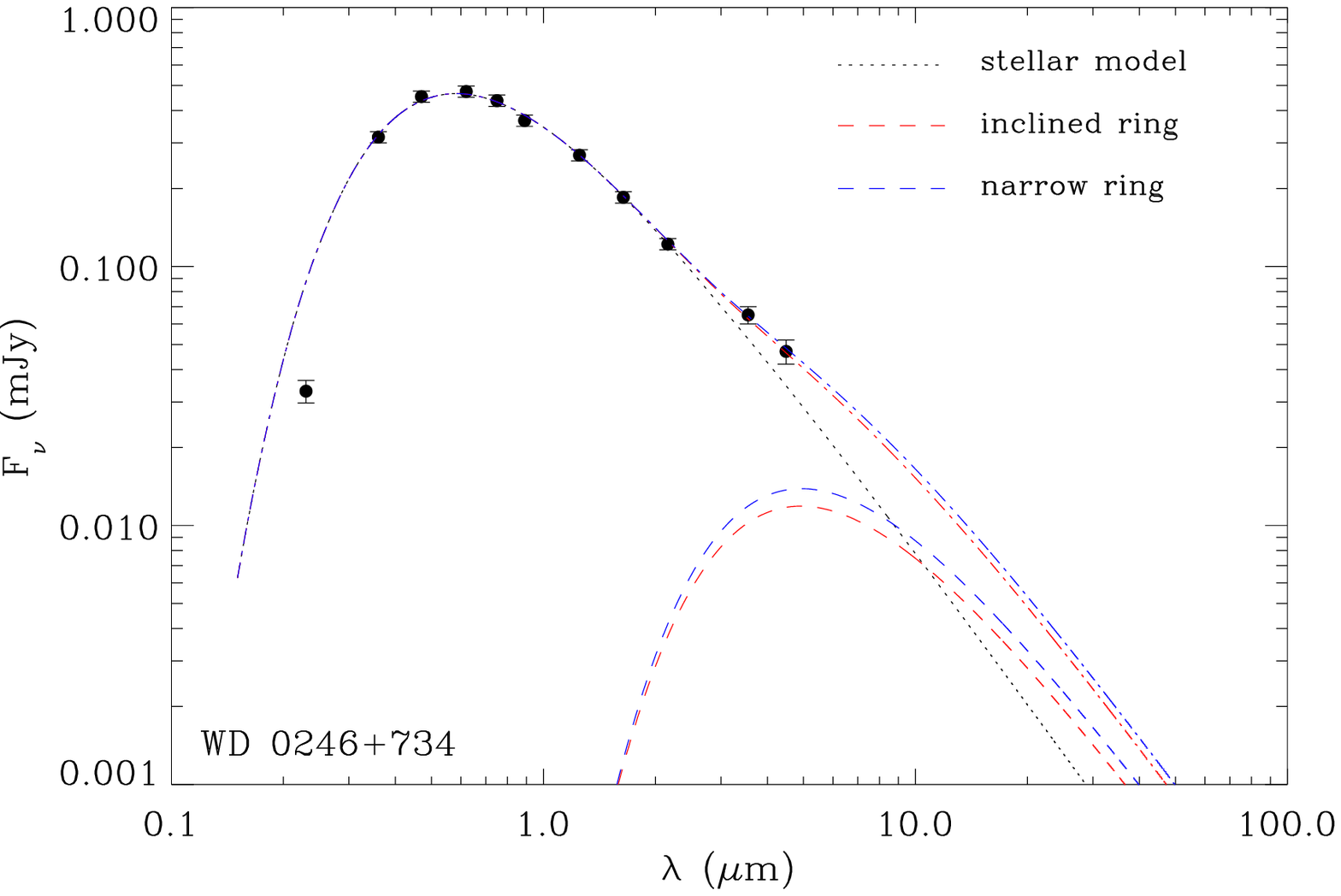}\\
  \includegraphics[width=8.5cm]{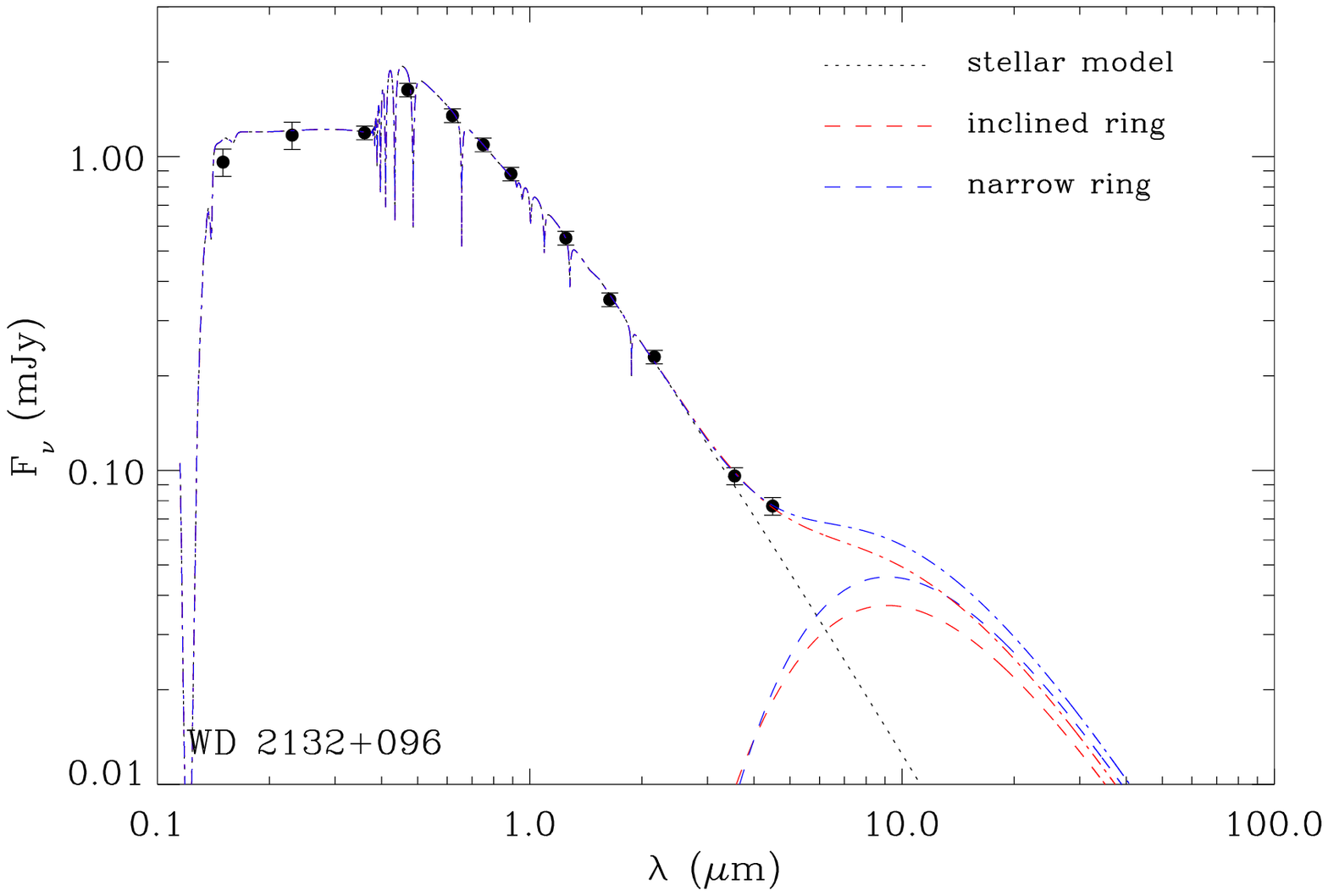}\\
  \end{tabular}
  \caption{(See Fig. 2 caption.) Disc model fits to the stars with infrared excess. The dashed red and blue lines show the disc emission model for highly inclined and narrow rings respectively (see Section 4.1 and Table \ref{disctable}), while the dash-dotted lines give the total emission from stellar atmosphere and disc combined.}
\label{excesses}
\end{figure}

This helium white dwarf shows traces of hydrogen in its spectrum together with Ca H \& K lines and is therefore classified as spectral type DBAZ \citep{WegnerSwanson1990, Bergeron2011}.
The excess in each of the $3.6$ and $4.5\mu$m wavelength bands is derived compared to a pure helium atmosphere white dwarf model of $T_{\rm eff}=15\,500$\,K. The star also shows a $\approx3\sigma$ excess in the $K_s$ band data obtained with SOFI at the NTT.
The white dwarf is isolated and there is no indication of photometric contamination. Fig. \ref{excesses} shows the disc spectral energy distribution for the two sets of models using the model input parameters listed in Table  \ref{disctable}.

\subsection{G245-58}
Using a pure helium atmosphere model, the optical and near-infrared photometry of this source is best fitted with $T_{\rm eff}=8250$\,K, and is shown in Fig. \ref{excesses}. 
Compared to this model, there is a $2.5\sigma$ infrared excess in both IRAC channels.
We adopt the weighted average flux obtained with \begin{small}DAOPHOT\end{small} and \begin{small}APEX\end{small} for the $3.6$ and $4.5\mu$m IRAC photometry (see Section 3.4 and Table \ref{Table_daophot_apex}). The $2.5\sigma$ excess is derived using conservative absolute calibration errors of 5 per cent. The likelihood that this is a real excess is strengthened in two ways. First, \citet{Reach2005b} report an absolute IRAC calibration error of less than 3 per cent, which, if adopted here, would increase the infrared excess to $>3\sigma$. Second, the excess is observed in both IRAC channels while the background source is more prominent at $4.5\mu$m.

Taken at face value, the best fitting disc model to the weak excess at G245-58 confines the circumstellar dust into a markedly narrow ring -- even for the high inclination model, the ring is smaller than four white dwarf radii. If the excess proves to be real, this is the first DZ star discovered with infrared excess, and the coolest with excess $3.6\mu$m emission.

We also fit metal-rich models of \citet{Dufour2007} to the short wavelength photometry of this source and derive a slightly cooler effective temperature of $T_{\rm eff}=7500$\,K and [Ca/He]=$-9.7$ (see Table \ref{targets}). These values are used in the following discussion and in Fig. \ref{mdotvteff} and \ref{histogram}.

\subsection{HS\,2132+0941}
We fit a $T_{\rm eff}=13\,000$\,K atmospheric model to optical and near infrared $JHK_s$ photometry for this hydrogen white dwarf. The star shows a $4\sigma$ infrared excess at $4.5\mu$m but no significant excess at shorter wavelengths (see Fig. \ref{excesses}). This is uncommon -- thermal disc emission is in general notable at white dwarfs at least in the $3.6\mu$m wavelength band and sometimes also in $K_s$. It is however not unique. Observations of the cool white dwarf G166-58 with $Spitzer$ IRAC revealed excess emission at $5.8$ and $7.9\mu$m only \citep{Farihi2008}, corresponding to dust no warmer than 400\,K. For PG\,1225-079, excess emission was observed only at 7.9$\mu$m \citep{Farihi2010b}. Since we only see excess flux in one wavelength band for HS\,2132+0941, the outer radius of the disc can not be well constrained. Until further observations are able to better constrain the nature of the excess, the best fitting disc model parameters are listed in Table \ref{disctable}.

\begin{table*}
\caption{Representative disc parameters obtained by fitting a model with high inclination or narrow radial extent.}
\centering
\begin{tabular}{l r r  r r r r l}
\hline\hline
WD	&	$T_{\rm eff}$	&	$R_{*}/d$	&	$T_{\rm in}$	&	$T_{\rm out}$	&	$r_{\rm in}$	&	$r_{\rm out}$	&	$i$\\	
	& ($K$)		&	($10^{-12}$)	&	($K$)	&	($K$)	&	($R_{\rm WD}$)	&	($R_{\rm WD}$)	&	(deg)\\
\hline
\multicolumn{8}{c}{Model 1: High inclination}\\
\hline
0246+734		&	8250		&	3.73	&	1200		&	900	&	7.8	&	11.4	&	80\\
1551+175		&	15500	&	1.80	&	1400		&	900	&	14.7	&	26.5	&	85\\
2132+096		&	13000	&	3.84	&	800		&	400	&	24.6	&	61.9	&	85\\
\hline
\multicolumn{8}{c}{Model 2: Narrow rings}\\
\hline
0246+734		&	8250		&	3.73	&	1000		&	970	&	9.9	&	10.4	&	60\\
1551+175		&	15500	&	1.80	&	1300		&	1200	&	16.2	&	18.1	&	60\\
2132+096		&	13000	&	3.84	&	600		&	520	&	36.0	&	43.6	&	60\\
\hline
\label{disctable}
\end{tabular}
\end{table*}

\subsection{WD\,0840--136}
This cool DZ star has an estimated temperature of $T_{\rm eff}\approx4900$\,K \citep{Giammichele2012}, while our coolest He atmosphere model has a temperature of $T_{\rm eff}=6000$\,K.  We therefore fit a blackbody function to the photometric measurements in $JHK_s$ from 2MASS and in \textit{VRI} from \citet{Subasavage2007}, finding decent agreement for $T=5750$\,K which is shown in Fig \ref{ModelFits3}.  This illustrates the well known behaviour of He atmospheric models with metals (and thus free electrons), which generally yield colours similar to a metal-free (or blackbody) model of higher temperature.

\section{Discussion}
\subsection{Disc fraction as a function of cooling age}
Metal polluted white dwarfs are common, yet only a fraction have observable disc emission. While the presence of metals in the atmospheres of helium white dwarfs may be explained as remnants of accretion from depleted discs due to the long diffusion times, the lack of discs around polluted hydrogen white dwarfs provides more of a puzzle.

\begin{figure}
 \centering
   \includegraphics[width=8.5cm]{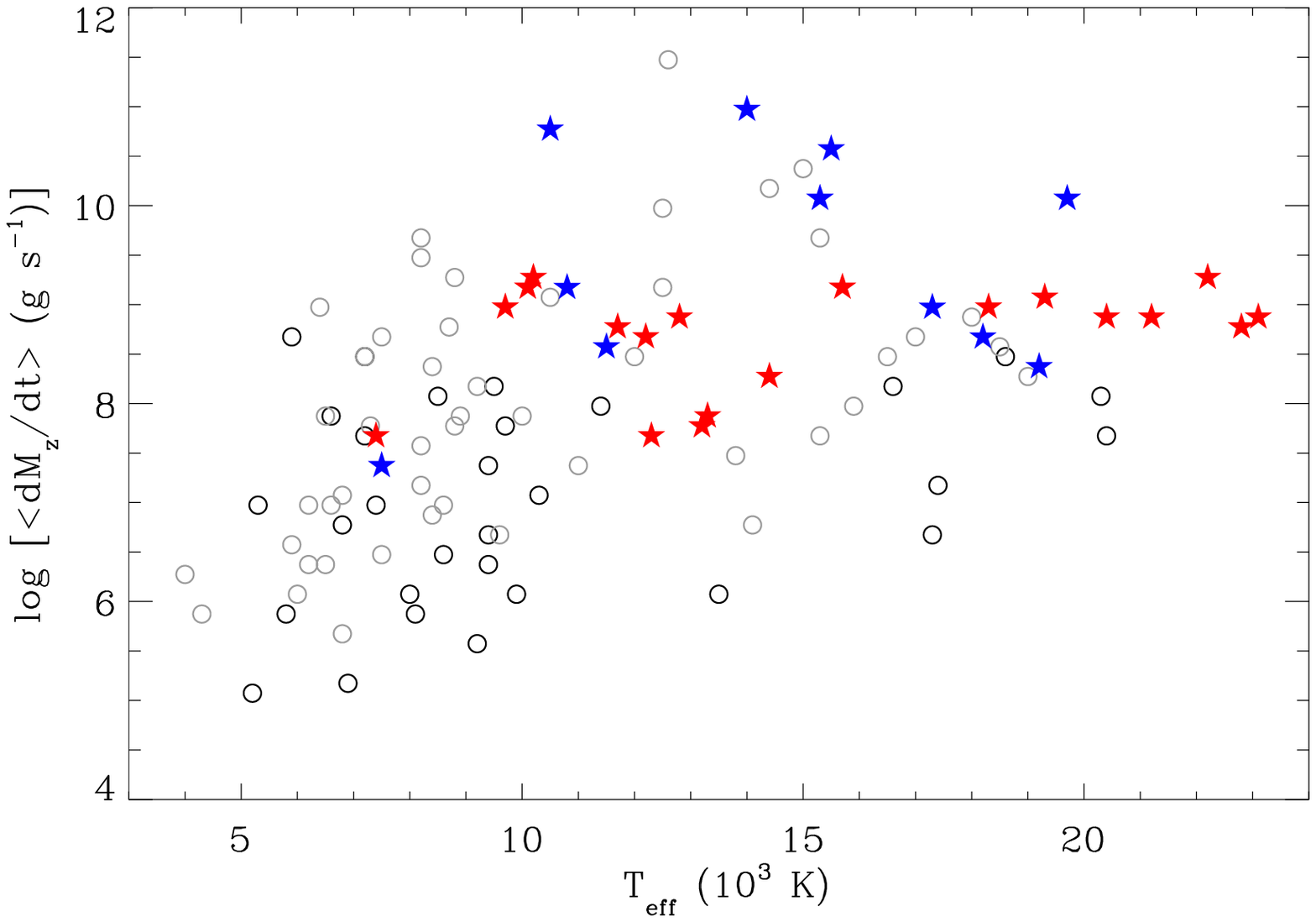}
    \caption{Mass accretion rate as a function of effective temperature for white dwarfs with known Ca abundance that have been observed with $Spitzer$ IRAC. Hydrogen white dwarfs with infrared excesses are plotted as red stars, and as black circles otherwise. Similarly, helium white dwarfs with infrared excess are plotted as blue stars, and as grey circles otherwise. }
\label{mdotvteff}
 \end{figure}

Previous studies have shown that the fraction of polluted white dwarfs with discs observable as infrared excess is related to both metal accretion rate and effective temperature (i.e. cooling age) of the stars \citep[e.g.][]{vonHippel2007,Farihi2009,XuJura2012}.
Fig. \ref{mdotvteff} depicts the time averaged mass accretion \citep{Farihi2012b, Girven2012} as a function of effective temperature for all polluted white dwarfs observed by $Spitzer$ IRAC. Adding our observations to already published data yield 104 stars, and corroborates trends noted in previous studies: infrared excess appears to be positively correlated with higher temperatures and younger ages, while there is a lack of detected excesses around cooler and older white dwarfs. However, because warmer white dwarfs have more opaque atmospheres than cooler ones, only the strongest metal absorption is detectable in their atmospheres from the ground, introducing a bias in the trend of accretion rate with temperature \citep{vonHippel2007, Zuckerman2010}. \citet{Koester2014} obtained spectra of 85 DA white dwarfs with the Cosmic Origins Spectrograph on the \textit{Hubble Space Telescope} and, when combined with ground-based observations, found no trend in accretion rate over cooling ages $\approx40$\,Myr to $\approx2$\,Gyr, or $T_{\rm eff}=5000-23\,000$\,K.

\begin{figure}
 \centering
   \includegraphics[width=6.8cm, angle=90]{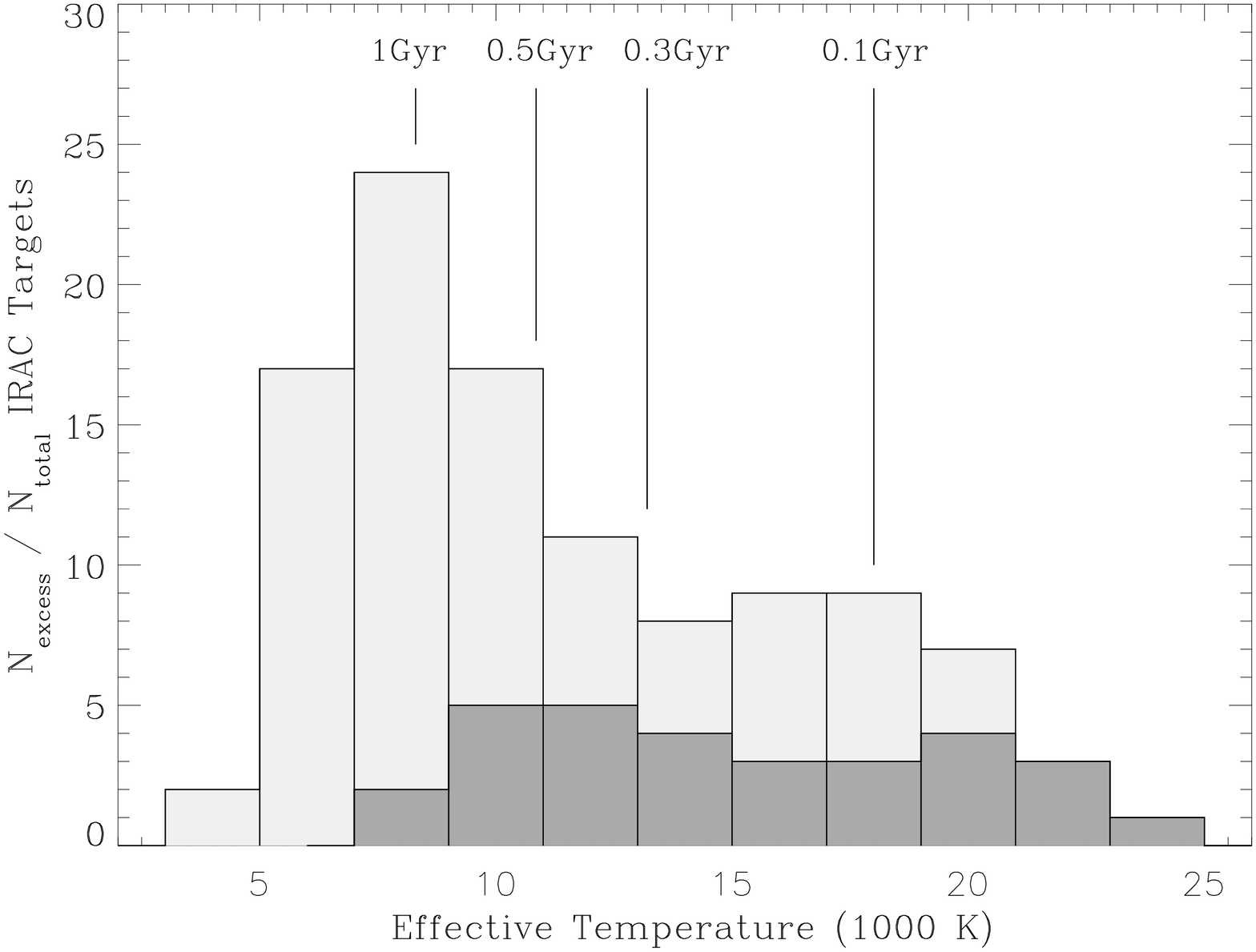}
    \caption{Fraction of polluted white dwarfs observed to have infrared excess with \textit{Spitzer} IRAC. Light grey bins represent the total number of observed white dwarfs, dark grey those with observed disc emission. While the fraction of polluted white dwarfs with discs is higher for warmer and younger white dwarfs, the discoveries in this study add to the number of cooler white dwarfs with faint excesses, suggesting there is a population of white dwarfs which still retain dust at old age.}
\label{histogram}
 \end{figure}

Fig. \ref{histogram} depicts the fraction of polluted white dwarfs that have been observed with \textit{Spitzer} IRAC for which disc emission was detected to those without. 
This distribution is somewhat biased since only the strongest pollution is detected for the warmest white dwarfs. High metal pollution is likely correlated with the most massive and therefore easily detected discs, thereby increasing the disc fraction for warm, young white dwarfs. 
Even so, from Fig. \ref{mdotvteff} it is clear that there is a trend of decreasing disc detections with decreasing temperature, even if considering only the most heavily polluted white dwarfs with [Ca/H(e)]$>-7.5$, consistent with previous findings \citep{Farihi2009, XuJura2012}. The fraction of metal enhanced white dwarfs with observable discs drops sharply around $T_{\rm eff}\approx10\,000$\,K, or around 500\,Myr, suggesting the disruption of large asteroids subsides on this timescale. 

\subsection{A few discs around old white dwarfs}
\citet{Jura2008} provided a model in which the lack of infrared excess emission among metal polluted white dwarfs is plausibly explained by accretion from tenuous or gaseous discs created as small asteroids enter a pre-existing disc at a non-zero inclination. In this model, erosion of the debris solids by sputtering creates a gaseous disc of vaporised asteroidal material, providing metals to be accreted by the host star but no observable infrared excess. Assuming a power-law mass distribution proportional to $M^{-2}$, similar to that of the solar system asteroid belt, implies that the majority of asteroids perturbed and destroyed within the tidal radius of the star are small, and massive discs may be expected to appear primarily for young systems in which larger asteroids (more massive than the pre-existing disc) are still relatively prevalent. In this picture, the observed trend of low detectable disc fraction for white dwarfs cooler than $T_{\rm eff}\la10\,000$\,K, supports a scenario where large planetesimal disruptions become rare. 

Before this study, only one white dwarf significantly cooler than $10\,000$\,K was known to have an infrared excess: the hydrogen white dwarf G166-58 \citep{Farihi2008}. The excess emission of this star is unusual in being detectable beginning at $6\mu$m. The possible discovery in this study of a weak infrared excess from a narrow dust ring around the cool DZ white dwarf G245-58, together with the subtle excess of G166-58, strengthens the idea that detectable discs become rare with age, but that there is a population of white dwarfs that still retain bright dust discs at ages of $\approx1$\,Gyr. The long metal diffusion time-scales for helium white dwarfs at this age imply that discs can prevail over time-scales of the order $10^6$ years or longer, consistent with the findings of \citet{Girven2012} and the model described in \citet{BochkarevRafikov2011}.

The fractional luminosity of discs around polluted white dwarfs has the potential to evolve with cooling age. \citet{Farihi2011} first noted that the highest $L_{\rm IR}/L_*$ occur only for relatively cool, dusty white dwarfs. 
This trend is consistent with an increasing area available to emitting solids interior to the fixed stellar Roche radius with decreasing stellar luminosity and temperature; the radius at which grains will rapidly sublimate moves inwards as a white dwarf ages (Rocchetto et al., in preparation). 
The relatively faint excesses around both G166-58 and G245-58 are remarkable in two ways. First, two detections among the coolest 42 stars (the three lowest $T_{\rm eff}$ bins in Fig. \ref{histogram}) clearly indicate a low frequency of infrared excess, especially considering these stars have nearly maximal areas available to emitting dust. Second, these two excesses are among the most subtle $L_{\rm IR}/L_*$ detected, and indicate that typically only tenuous rings exist at these ages.

\subsection{Subtle discs/narrow rings are common}
Prior studies suggested that narrow rings may be present around metal polluted stars without observed infrared excess \citep{Farihi2010b}, and the three new discs discovered here corroborate this view. In fact, only \textit{Spitzer} observations were capable of confirming dust in these particular systems; strong infrared excesses would have been previously detected in the $K$-band from the ground \citep{Kilic2006}, or with
\textit{WISE} \citep{Hoard2013}. Owing to the degeneracy between the spatial extent and inclination for faint discs such as these, we provided two representative disc models: one narrow at moderate inclination and the other more extended but at high inclination. However, it is geometrically unlikely that all three of these discs have high inclinations, and therefore the subtle excesses imply that the disc material is confined to narrow rings around each star. This is also likely for previously published infrared excesses with low $L_{\rm IR}/L_*$.

The disc model for G245-58 essentially represents the limit of what is confidently detectable using \textit{Spitzer} IRAC. While 0.5 stellar radii ($1R_{\rm WD}\approx8800$\,km) is the most narrow ring inferred for any polluted white dwarf, it is still comparable to many planetary rings \citep{Esposito1993}; e.g. the main ring of Jupiter is 6500\,km wide. Dust rings as narrow as a few tenths of an Earth radius are suggested by the collective infrared properties of polluted white dwarfs, but are undetectable with current instrumentation. Our data support the existence of a population of white dwarfs with very faint dust discs, most of which are currently beyond detection limits and implying that dust discs are more common than what is observed. Future missions like the \textit{James Webb Space Telescope} (\textit{JWST}) and the \textit{Space Infrared telescope for Cosmology and Astrophysics} (\textit{SPICA}) with infrared spectroscopic capabilities should detect many more faint discs.
 
 \section{Summary and conclusions}
 We observed 15 metal polluted white dwarfs with \textit{Spitzer} IRAC to search for infrared excess emission indicative of circumstellar dust. Such excesses were found for two or possibly three white dwarfs in our target sample: KUV\,15519+1730, HS\,2132+0941 and G245-58. Notably, all the dust emissions are subtle in $L_{\rm IR}/L_*$; such subtle excesses not only imply narrow rings around these three stars, but also for faint discs discovered in previous studies. All together these data suggest an undetected population of weakly emitting discs at polluted white dwarfs, possibly resembling the planetary rings of the outer solar system.
 
 The detection of discs at 1\,Gyr white dwarfs supports previous findings that discs can persist for timescales of Myr or longer. For the infrared bright discs so far detected, the frequency is highest for warm and young white dwarfs, and clearly declines on a timescale of $500$\,Myr, possibly implying that the reservoir of large asteroids further out in the system is depleted on this timescale.

\section*{Acknowledgments}
JF gratefully acknowledges the support of the STFC via an Ernest Rutherford Fellowship. CB and MR acknowledge support from the STFC.
We are grateful to D. Koester for providing the white dwarf atmospheric models used.
Balmer/Lyman lines in the models were calculated with the modified 
Stark broadening profiles of \citet{TremblayBergeron2009}, kindly made available by the authors. The authors thank the anonymous reviewer for suggestions that improved the manuscript.
This work is based on observations made with the Spitzer Space Telescope, which is operated by JPL/Caltech under a contract with NASA. This work has made use of data products from the SDSS, which is managed by the Astrophysical Research Consortium for the Participating Institutions, and from 2MASS, which is a joint project of the University of Massachusetts and IPAC/Caltech funded by NASA and NSF. 
This research has made use of the VizieR catalogue access tool and the SIMBAD database operated at CDS, Strasbourg, France, and the SAO/NASA Astrophysics Data System.

\label{lastpage}

\end{document}